\documentstyle[prb, aps, epsf]{revtex}
\begin{document}
\title{Perturbation calculations on interlayer transmission rates  from symmetric to antisymmetric channels in parallel armchair nanotube junctions}

\author{Ryo Tamura}
\address{Faculty of Engineering, Shizuoka University, 3-5-1 Johoku, 
  Hamamatsu 432-8561, Japan}

\maketitle

\begin{abstract}
 Partially overlapping two parallel armchair nanotubes  
 are investigated theoretically
 with the $\pi$ orbital tight bonding  model.
 Considering the interlayer Hamiltonian  as perturbation,
 we obtain approximate analytical formulas of
 the interlayer transmission rates $T_{\sigma',\sigma}$
 from channel $\sigma$ to $\sigma'$ 
 for all the four combinations $(\sigma',\sigma)=(\pm,\pm)$ and $(\pm,\mp)$,
 where  suffixes  $+$ and $-$ represent symmetric and antisymmetric channels,
 respectively,  with respect to the mirror plane 
 of each tube.
Landauer's formula conductance  is equal to the sum of them
 in units of  $2e^2/h$.
According to the perturbation calculation,
 the interlayer Hamiltonian is transformed into the parameter
 $w_{\sigma',\sigma}$ that determines the analytical formula
 of $T_{\sigma',\sigma}$.
 By comparison with the exact numerical results,
  the effective range of the analytical formulas is discussed.
 In the telescoped coaxial contact,
 the off-diagonal part $T_{-,+}+T_{+,-}$ is
 very small compared to the diagonal part $T_{+,+}+T_{-,-}$.
 In the side contact, on the other hand,
 the off-diagonal part
 is more significant than the diagonal part
 in the zero energy peak of the conductance.

\end{abstract}

\twocolumn

\section{introduction}
 In the growing area of carbon nanotubes (NT) \cite{NT-book,NT-review}
 and graphenes (GR) \cite{graphene-review}, 
 interlayer interaction
 has important roles.
 In the NT system, it brings about  pseudogaps \cite{pseudo-gap}, nearly free electron states \cite{3-NFE}, and formation of single wall NT ropes \cite{4-rope}.
In the multi-layer GR, it causes 
 band gaps under the electric field \cite{bilayer-E-fierld},
 and superconductivity of twisted bilayer GR \cite{6-twist}.
The two inequivalent Fermi points K and K' of the single layer
 are called  valleys.
Effective mass theory shows that a boundary between monolayer and bilayer GR
works as valley current filters \cite{17-mono-double-junction}.
Since interlayer bonds  are  much  weaker than intralayer bonds,  interlayer sliding and rotation occur keeping the honeycomb lattice. 
Telescopic extension of multiwall NTs has been
 investigated experimentally \cite{experiment-telescope-mechanical} and theoretically \cite{theory-telescope-mechanical} as GHz oscillators and  nano springs.
Interlayer interaction energy and force
 were calculated for  a stack of GR flakes 
 \cite{gra-gra-MD} and for a NT on a GR layer \cite{NT-graphene-force-theory}.
Molecular dynamic calculations
 indicate that AB stacking is the
 most stable in the NT-GR connection \cite{gra-NT-AB}.
The interlayer force
 is usually classified 
 to van der Waals force caused by virtual dipole-dipole interaction
 that could exist without the interlayer orbital
 overlap \cite{JJ-Sakurai}.
 The electronic structures, however, are
 described well by the tight binding (TB) model
 with the interlayer transfer integrals
 that originate from  the interlayer orbital overlap \cite{gra-band}.
 In the present paper,  the interlayer transfer integral
 is termed the interlayer bond. 
Interlayer 'covalent' bonds 
induced  by beam irradiation, heating and defects \cite{10-covalent}  are excluded in our discussion as they  hinder the nearly free interlayer motion.

 Among various multi-layer systems,
  a single layer $\downarrow$
  partially  overlapping with another single
 layer  $\uparrow$ is outstanding in the
  relation between the interlayer bonds 
 and the conductance.
It is represented by 
(L,$\downarrow$)-(D,$\downarrow$,$\uparrow$)-(R,$\uparrow$)
 where interlayer bonds are limited to the overlapped region D.
 Connecting  the source and drain electrodes  to single layer
 regions L and R, respectively,  we can force the net
 current to flow through the interlayer bonds.
In contrast to this $\downarrow$-$\uparrow$ junction,
 the net  current between $\downarrow$ and $\uparrow$ is zero 
 in the junctions (L,$\downarrow$)-(D,$\downarrow$,$\uparrow$)-(R,$\downarrow$) where both the source and drain electrodes are connected to
 $\downarrow$ \cite{gra-gra-10-LDL}.
 The $\downarrow$-$\uparrow$ conductance was measured for the telescoped
 NTs \cite{experiment-telescope-conductance}.
 The Landauer's formula conductance 
 of $\downarrow$-$\uparrow$ junctions
 has been reported.
 The combinations $\downarrow$- $\uparrow$ are  GR-GR \cite{gra-gra-tele-cond}, NT-GR \cite{15-gra-NT}  and NT-NT.
Telescoped coaxial contacts \cite{theory-telescope-conductance,previous-paper,ref-2,sawai,Uryu}  and side contacts \cite{10-side-contact,10-side-contact-ref-1} 
were discussed  for the NT-NT junctions.
Comparisons between the two contacts were also reported \cite{side-tele,side-tele-perturb}.

 The Landauer's formula conductance is the sum of  the interlayer
 transmission rates $T_{\sigma',\sigma}$  
 of which indexes $\sigma'$ and $\sigma$ denote
 channels of R  and L, respectively.
Wave numbers $k_1$ and $k_2$  of region D appear
  in  the dependence of $T_{\sigma',\sigma}$  on the  overlapped length  as  the periods of the beating,
  $2\pi/|k_1-k_2|$ and   $2\pi/|k_1+k_2|$.
 In addition to this $(k_1,k_2)$ characteristic,
 we can show that $T_{\sigma',\sigma}$ is proportional to $|W|^2$
  considering the interlayer bond  $W$  as perturbation \cite{previous-paper,Uryu,side-tele-perturb}.
 It is termed the $|W|^2$ characteristic here.
 The $(k_1,k_2)$ and  $|W|^2$ characteristics appear in the period
 and in the amplitude of the oscillation, respectively,
 while both originate from  $W$.
 Whereas the numerical calculation method
 about $T_{\sigma',\sigma}$ 
 has been established \cite{Datta},
  it does not diminish the value of the perturbation calculation
 producing analytical formulas.
 Without the perturbation calculation, one might assume an analytical formula
 of which fitting parameters  are optimized  for
 the coincidence with the numerical
 results.
In this fitting method, however,
 the fear is that choice of the formula may become
 arbitrary.
When  we know the exact eigen states of the unperturbed Hamiltonian, however,
 we can derive the unique perturbation expansion \cite{JJ-Sakurai}.

In the present paper,  $\downarrow$
 and $\uparrow$  are chosen to be
  parallel $(n_\downarrow,n_\downarrow)$
 and $(n_\uparrow,n_\uparrow)$ armchair NTs,
  because their mirror symmetry and small unit cell enable us
 to perform
 the analytical perturbation  calculation.
 Figure 1 shows the (a) side contact
 and (b) telescoped coaxial contact.
 The mirror symmetry of each NT
 is indicated  by  $\sigma=+$ and $\sigma=-$ 
 in the suffixes of $T_{\sigma',\sigma}$.
 The $(k_1,k_2)$ characteristic does not appear  in the nonparallel crossed NT junction  without  periodicity in region D \cite{X-junction-cond}.
 In the chiral NT junctions,  the large unit cell of  region D makes
 the $(k_1,k_2)$ characteristic complicated \cite{Uryu,side-tele-perturb}.
 In the reported theoretical works
 on the $(n_\uparrow,n_\uparrow)$-$(n_\downarrow,n_\downarrow)$ junctions,
 the diagonal transmission rates $T_{+,+}$
 and $T_{-,-}$ and the sum $\sum_{\sigma}\sum_{\sigma'}T_{\sigma',\sigma}$ have been discussed, but
 the off-diagonal transmission rates $T_{+,-}$
 and $T_{-,+}$ have been neglected.
In this paper, we derive the analytical formulas of all  four
 $T_{\sigma',\sigma}$ and show how the 
 $|W|^2$  and   $(k_1,k_2)$ characteristics appear there.

\section{geometrical structure and tight binding model}\label{sec-TB}
As is shown by Fig. 1,  the tube axis of $\downarrow$ is chosen to be
 $z$ axis.
The atomic $z$ coordinates in tubes  $\downarrow$ and $\uparrow$
 are $aj/2$ and $aj/2+\Delta z$, respectively, with integers $j$, the lattice constant $a=0.246$ nm and a small translation $|\Delta z| < a/4$.
The atomic $y$ coordinates  of tube $\xi (= \downarrow, \uparrow)$
 are represented by
 $R_\xi\sin\theta^{\xi}_{j,i}$
 with the tube radius $R_\xi=\frac{\sqrt{3}a}{2\pi}n_\xi$.
 The  angles
 $\theta^{\downarrow}_{j,i}=\frac{\chi_{j,i}}{n_\downarrow}$
and $\theta^{\uparrow}_{j,i}=\frac{\chi_{j,i}}{n_\uparrow}-\frac{2\pi}{3n_\uparrow}+\Delta \theta$  are measured in the opposite direction 
with  positive integers $i$, $\chi_{j,i}\equiv \pi(i-\frac{(-1)^i}{6}-\frac{(-1)^j}{2})$ and a small rotation $|\Delta\theta| <\pi/n_\uparrow$.
Thus the atomic $x$ coordinates are
 $R_\downarrow\cos\theta^{\downarrow}_{j,i}$  for   tube $\downarrow$, and
 $D+R_\uparrow+R_\downarrow-R_\uparrow\cos\theta^{\uparrow}_{j,i}$ for  tube $\uparrow$. Here
 $D= $ 0.31 nm is the interlayer distance for the side-contact while
$D=-R_{\downarrow}-R_{\uparrow}$ for the coaxial contact.
The former is the same as Ref. \cite{side-tele}.
When $|n_\downarrow - n_\uparrow| =5$,
 the interlayer distance of the coaxial contact
 is close to that of graphite. 
For example, Fig. 2 shows
 the interlayer configuration
 in the case where $(n_\downarrow, n_\uparrow, \Delta\theta,\Delta z)=
(10,15,0,0)$.
Tubes  $\downarrow$ and  $\uparrow$
 have 'AB' and 'ab' sublattices where odd $i$ sites correspond to 'A' and 'a' sublattices. In Fig. 2(a) for the side contact, 1A and 1a (2B and 2b)
 sites correspond to $i=1$ ($i=2$).
The interlayer configuration
  in the side contact  is similar to the  Ab stacking of the bilayer GR
 when $(\Delta\theta,\Delta z)=(0,0), (\frac{-2\pi}{3n_\uparrow},0)$.

The $\pi$ orbital TB equations with energy $E$  in region 
 D are represented by
\begin{equation}
E\vec{c}_j^{\;\rm (D)}=\sum_{\Delta j=-1}^1 H^{(j,\Delta j)}\vec{c}_{j+\Delta j}^{\;\rm (D)}
\label{recursion}
\end{equation}
 where $\;^t\vec{c}^{\;(\rm D)}_j=(\;^t\vec{c}^{\;(\rm D,\downarrow)}_j,
\;^t\vec{c}^{(\;\rm D,\uparrow)}_j)$ .
The matrix $H^{(j,\Delta j)}$ is partitioned  as
\begin{equation}
H^{(j,\Delta j)}=
\left(
\begin{array}{cc}
h_{\downarrow}^{(j,\Delta j)}
, &  
W^{(j,\Delta j)}
\\

\;^tW^{(j+\Delta j,-\Delta j)}, & h_{\uparrow}^{(j,\Delta j)}
\end{array}
\right).
\label{def-H}
\end{equation}
The blocks $h$ and $W$ correspond to  intralayer and interlayer elements, respectively.
 Figure 3 shows a schematic diagram  of the tight binding Hamiltonian.
 As $H^{(j,\Delta j)}$ is  the block  of the Hamiltonian matrix  partitioned 
by the {\it half} lattice constant $a/2$, $H^{(j+2,\Delta j)}=H^{(j,\Delta j)}$.
 The $(i,i')$  element of $W^{(j,\Delta j)}$
 is  defined by 
$ t_1e^{\frac{d-r}{L_c}}\Theta(r-r_c)|\cos\phi|$ 
 where $\phi=\theta_{j,i}^{\downarrow}+\theta_{j+\Delta j,i'}^{\uparrow}$, $t_1=$ 0.36 eV, $d= $ 0.334 nm, 
$L_c=$ 0.045 nm, 
 the cut-off radius $r_c =0.39$ nm, $r$ denotes the atomic distance and $\Theta$ is the step function defined by 
$\Theta(x)=1$ for
 negative $x$ and
$\Theta(x)=0$ for
 positive $x$.
The elements between nearest neighbors 
 are $h^{(j,0)}_{\xi,2m-1,2m}$, $h^{(j,0)}_{\xi,2m,2m-1}$, $h^{(1,\pm 1)}_{\xi,2n_\xi,1}$, $h^{(1,\pm 1)}_{\xi,m-1,m}$, $h^{(2,\pm 1)}_{\xi,1,2n_\xi}$
 and $h^{(2,\pm 1)}_{\xi,m,m-1}$  with integers $m$.
 They  are equal to the negative constant $-t = -2.75$ eV 
 while the other elements of $h^{(j,\Delta j)}_\xi$ are zero.
 Since $h_\xi^{(1,0)}=h_\xi^{(2,0)}$ and $h_\xi^{(j,1)}=h_\xi^{(j,-1)}$, 
  we use the abbreviation $h_\xi^{(0)}$ and $h_\xi^{(1)}$ in Fig. 3.
On the other hand,  relations $W^{(1,0)}=W^{(2,0)}$ and $W^{(j,1)}=W^{(j,-1)}$ 
 do not generally hold. The latter relation $W^{(j,1)}=W^{(
j,-1)}$ is
 valid only when $\Delta z =0$.

 Our calculation and Refs. \cite{sawai,Lambin}
 are the same  in the TB model except that $t_1$  has
 two values 0.36 eV and 0.16 eV in Refs. \cite{sawai,Lambin}.
As this multivalued $t_1$ model
 was derived from first principle calculation data on 
   multiwall NTs, it
 may not be effective for the side contact.
In our calculation,  $t_1$ is fixed at the single value 0.36 eV 
 and the  geometrical structure 
  is simplified compared to the actual one
  as a first guess.

\section{Method of calculation} 
In order to obtain the transmission rate, we calculate the 
 scattering matrix ($S$ matrix).
The $S$ matrix has two useful characteristics.
Firstly,  unitarity  $\;^tS^*=S^{-1}$ is guaranteed by conservation
 of the probability.
When there is time reversal symmetry, $\;^tS =S$ also holds.
These symmetries  proved in Appendix A can be used as  verification of  the obtained  results.
Secondly, 
$S$ matrix is directly
 related to the ratio between incident and scattered wave amplitudes.
It leads us to  an intuitive formula showing that
multiple reflection between the two boundaries
 causes  the transmitted wave.

\subsection{exact numerical calculation}\label{sec-exact}

Equation (\ref{recursion}) enables us to obtain the transfer matrix $\Gamma^{(\rm D)}$ that satisfies
$(\;^t\vec{c}^{\;(\rm D)}_{2m+1},\;^t\vec{c}^{\;(\rm D)}_{2m+2})
=(\;^t\vec{c}^{\;(\rm D)}_{2m-1},\;^t\vec{c}^{\;(\rm D)}_{2m})\;^t\Gamma^{(\rm D)}$.
Replacing $W^{(j,\Delta j)}$ with zero, we also obtain
 the transfer matrices $\Gamma^{(\rm L)}$ 
 and  $\Gamma^{(\rm R)}$  for regions L and R.
With   a set of  linearly independent eigen vectors 
$\vec{u}^{\;(\mu)}_{l}$ 
satisfying
$\Gamma^{(\mu)}\vec{u}_l^{\;(\mu)}=\lambda_l^{(\mu)}\vec{u}_l^{\;(\mu)}$ ,
 we can expand $\vec{c}_j^{\;(\mu)}$ as
\begin{equation}
 \left(
\begin{array}{c}
\vec{c}^{\;(\mu)}_{2m-1} \\
\vec{c}^{\;(\mu)}_{2m} \\
\end{array}
\right)
=\sum_{l=-2n_\mu}^{2n_\mu}
\vec{u}^{\;(\mu)}_l
\left(\lambda^{(\mu)}_l
\right)^m
\gamma_l^{(\mu)}
\label{f-expand}
\end{equation}
where   $l \neq 0$,  $\lambda_{-l}^{(\mu)}=1/\lambda_l^{(\mu)}$,  $n_{\rm L}=n_{\downarrow},n_{\rm R}=n_{\uparrow}$ and  $ n_{\rm D}=n_{\rm L}+n_{\rm R}$.
The eigen vectors are ordered according  to the following rules  (i) for 
 propagating waves and (ii) for evanescent waves.
Here $\overline{N}_\mu$ denotes the channel number of region $\mu$.
(i)When $1 \leq l \leq \overline{N}_\mu$
,  $\left|\lambda_l^{(\mu)}\right|=1$, $\vec{u}_{-l}^{\;(\mu)}=\left(\vec{u}_{l}^{\;(\mu)}\right)^*$  and the probability flow of $\vec{u}_l^{\;(\mu)}$ is positive.
Note that $|\vec{u}_l|^2 \neq 1$. The normalizatiion of $\vec{u}_l^{\;(\mu)}$ is defined by Appendix A. 
(ii)When $\overline{N}_\mu+1 \leq l \leq 2n_\mu$,  
$\left|\lambda_l^{(\mu)}\right|< 1$.

The boundary conditions for the LD junction
 are
\begin{eqnarray}
\left(
\begin{array}{c}
\vec{c}^{\;(\rm L)}_{j_{\rm l}+1} 
\\
\vec{c}^{\;(\rm L)}_{j_{\rm l}}
\\
0
\end{array}
\right)
=
\left(
\begin{array}{c}
\vec{c}^{\;(\rm D,\downarrow)}_{j_{\rm l}+1} 
\\
\vec{c}^{\;(\rm D,\downarrow)}_{j_{\rm l}}
\\
\vec{c}^{\;(\rm D,\uparrow)}_{j_{\rm l}}
\end{array}
\right)
+
\left(
\begin{array}{c}
\frac{1}{h_\downarrow^{(j_{\rm l} ,1)}}W^{(j_{\rm l} ,1)} \vec{c}^{\;(\rm D,\uparrow)}_{j_{\rm l}+1}
\\
0
\\
0
\end{array}
\right)
\label{LD}
\end{eqnarray}
and those of the DR junction are
\begin{eqnarray}
\left(
\begin{array}{c}
\vec{c}^{\;(\rm R)}_{j_{\rm r}}
\\
\vec{c}^{\;(\rm R)}_{j_{\rm r}+1}
\\
0 
\end{array}
 \right)
=
\left(
\begin{array}{c}
\vec{c}^{\;(\rm D,\uparrow)}_{j_{\rm r}}
\\
\vec{c}^{\;(\rm D,\uparrow)}_{j_{\rm r}+1}
\\
\vec{c}^{\;(\rm D,\downarrow)}_{j_{\rm r}+1}
 \\
\end{array}
 \right)
+
\left(
\begin{array}{c}
\frac{1}{h_\uparrow^{(j_{\rm r}+1 ,1)}}
\;^tW^{(j_{\rm r} ,1)} \vec{c}^{\;(\rm D,\downarrow)}_{j_{\rm r}}
\\
0 
\\
0
\end{array}
 \right)
\label{DR}
\end{eqnarray}
 where  $j_{\rm l}$ and  $j_{\rm r}$ denote $j$ at the boundaries
 as is shown by Fig. 1.
 The geometrical overlapped length equals  $z_{\rm R}-z_{\rm L}=-\Delta z+ (j_{\rm r}-j_{\rm l} -1)a/2$.
Without losing generality,  $j_{\rm l}$ is either $-1$ or 0.
Derivation of Eqs. (\ref{LD}) and (\ref{DR}) is shown by Appendix B.
Since Eq. (\ref{f-expand}) must not diverge at $j = \pm \infty$, 
 $\gamma_l^{(\rm L)}=0 $ and $\gamma_{-l'}^{(\rm R)}=0 $
 when $l > \overline{N}_{\rm  L}$ and $l' > \overline{N}_{\rm R}$.
Thus the number of nonzero variables  
is $M_{\rm var}=2n_{\rm L}+2n_{\rm R}+\overline{N}_{\rm L}+\overline{N}_{\rm R}+4n_{\rm D}$.
On the other hand, the number of conditions  
 is $M_{\rm cond}=2n_{\rm L}+2n_{\rm R}+4n_{\rm D}$
 to which contributions of  Eqs. (\ref{LD}) and (\ref{DR})
 are  $ 4n_{\rm L}+2n_{\rm R} $ and $ 4n_{\rm R}+2n_{\rm L} $ , respectively.
Accordingly the number of independent variables is $M_{\rm var}-M_{\rm cond}=\overline{N}_{\rm L} +\overline{N}_{\rm R}$.
Choosing $\;^t\vec{\gamma}_+^{\;(\rm L')}=(
\gamma_{1}^{(\rm L)},\gamma_{2}^{\;(\rm L)},\cdots ,
\gamma_{ \overline{N}_{\rm L}}^{(\rm L)})$
and
$\;^t\vec{\gamma}_-^{\;(\rm R')}=(
\gamma_{-1}^{(\rm R)},\gamma_{-2}^{\;(\rm R)},\cdots ,
\gamma_{ -\overline{N}_{\rm R}}^{(\rm R)})$
for the independent variables, 
 we obtain the scattering matrix $S_{\rm RL}$ satisfying
\begin{equation}
\left(
\begin{array}{c}
\vec{\gamma}^{\;(\rm L')}_-
\\
\vec{\gamma}^{\;(\rm R')}_+
\end{array}
\right)
=
\left(
\begin{array}{cc}
r_{\rm LL}, & t_{\rm LR} 
\\
t_{\rm RL}, & r_{\rm RR}\\
\end{array}
\right)
\left(
\begin{array}{c}
\vec{\gamma}^{\;(\rm L')}_+
\\
\vec{\gamma}^{\;(\rm R')}_-
\end{array}
\right)
\label{smatrix}
\end{equation}
where $S_{\rm RL}$ is partitioned into reflection blocks 
$r_{\rm LL},  r_{\rm RR} $ and 
transmission blocks $t_{\rm LR},  t_{\rm RL} $.
Detail of the numerical calculation is shown by Appendix B.
The energy $E$ we consider here is close to zero so that
 $\overline{N}_{\rm L}=\overline{N}_{\rm R}
=2$.

 \subsection{approximate analytical calculation}\label{sec-appro}
We consider  the Bloch state $(^t\vec{c}^{\;\rm(D)}_{2m-1},^t\vec{c}^{\;\rm(D)}_{2m})=e^{ikam}\;^t\vec{b}$ for the periodic system corresponding to region D.
Equation (\ref{recursion}) is transformed into
 the  eigen value equation $E_l\vec{b}_l=H(k)\vec{b}_l$ 
 with the Hamiltonian
\begin{eqnarray}
H(k)&=& 
\left(
\begin{array}{cc}
H^{(1,0)},
 & 
H^{(1,1)}
\\
H^{(2,-1)}
,&
H^{(2,0)} 
\end{array}
\right)
\nonumber \\
&& +
\left(
\begin{array}{cc}
0,
 & 
e^{-ika}H^{(1,-1)}
\\
e^{ika}H^{(2,1)}
,&
0
\end{array}
\right).
\label{Hk}
\end{eqnarray}
In the perturbation calculation, $H(k)=H_0(k)+\beta V(k)$ where
 $H_0(k)$ and $\beta V(k) $ correspond to intralayer $h_{\uparrow,\downarrow}^{(j,\Delta j)}$
 and  interlayer $W^{(j,\Delta j)}$, respectively.
The constant $\beta=1$
is introduced for counting the times the perturbation $V$ 
 enters, namely, $E_l$ and $\vec{b}_l$ 
are expanded  as
$\vec{b}_l=\vec{b}_l^{\;[0]}+\beta\vec{b}_l^{\;[1]}+\beta^2\vec{b}_l^{\;[2]}+\cdots$
 and 
$E_l=E_l^{[0]}+\beta E_l^{[1]}+\beta^2E_l^{[2]}+\cdots$.
We choose  the unperturbed states near zero energy, 
\begin{equation}
E_{\sigma,\tau}^{[0]}
=\sigma t\left (2\cos\frac{ka}{2} -1 \right) 
\label{El0}
\end{equation}
\begin{equation}
\vec{b}_{\sigma,\tau}^{\;[0](\zeta)}=
\left(
\begin{array}{c}
\vec{d}_{\sigma,\tau}^{\;[0](\zeta)} \\
\exp\left(i\frac{k}{2}a+i\pi \right)\vec{d}_{\sigma,\tau}^{\;[0](\zeta)}
\end{array}
\right)
\label{e-d}
\end{equation}
 where
\begin{equation}
\;^t\vec{d}_{\sigma,\tau}^{\;[0](\zeta)}
=(\;^t\vec{g}_{\downarrow,\sigma},\;\tau f_\sigma^{(\zeta)}\;^t\vec{g}_{\uparrow,\sigma})
\label{def-dzero}
\end{equation}
\begin{equation}
\;^t\vec{g}_{\xi,\sigma}=
\frac{1}{\sqrt{8n_\xi}}
(1,\sigma,1,\sigma,\cdots,1,\sigma)
\label{def-g}
\end{equation}
with a constant factor $f_\sigma^{(\zeta)}$. 
The auxiliary index $\zeta =\pm$ indicates
 that the wave number $k$ is close to $\zeta 2\pi/(3a)$.
 Relation of $\vec{b}, \vec{d}, \vec{g}$ to
 notation of Sec.\ref{sec-exact} is illustrated 
 by Fig. 4.
In Eqs. (\ref{El0}) and (\ref{e-d}),  index $l$ 
 is replaced by  $(\sigma,\tau)=(+,+),(-,+),(+,-),(-,-)$ 
 where $\sigma$ indicates the mirror symmetry
 of the isolated tubes.
Since we consider energy region $|E| \ll t$ and the Brillouin zone $|ka| \leq \pi$,  the phase $\pi$ of Eq. (\ref{e-d})
 is necessary.
If we deleted the phase $\pi$ of Eq. (\ref{e-d}),
 Eq. (\ref{El0}) would be  changed into $ E_l^{[0]}
=-\sigma t\left (2\cos\frac{ka}{2} +1 \right) $.
In this notation, the wave number $k$ at zero energy  would be  $\pm 4\pi/(3a)$ outside the Brillouin zone $|k|\leq \pi/a$.

The matrix element of the perturbation
$V_{(\sigma',\tau'|\sigma,\tau)}^{(\zeta)}= 
\;^t\left(\vec{b}_{\sigma',\tau'}^{\;[0](\zeta) }\right)^* V\left (\zeta \frac{2\pi}{3a} \right) \vec{b}_{\sigma,\tau}^{\;[0](\zeta)}
$ 
is represented by
\begin{equation}
V_{(\sigma',\tau'|\sigma,\tau)}^{(\zeta)}= 
\tau f_\sigma^{(\zeta)}w_{\sigma,\sigma'}^{(\zeta)}
+\tau'\left( f_{\sigma'}^{(\zeta)}w_{\sigma',\sigma}^{(\zeta)} \right)^*
\label{d1}
\end{equation}
where $k$ is approximated by $\zeta 2\pi/(3a)$,
\begin{equation}
w_{\sigma',\sigma}^{(\zeta)}
=\eta_{\rm A,a}^{(\zeta)}+\sigma\sigma'\eta_{\rm B,b}^{(\zeta)}
+\sigma'\eta_{\rm A,b}^{(\zeta)}+\sigma\eta_{\rm B,a}^{(\zeta)}
\label{overline-w}
\end{equation}
\begin{equation}
\eta_{s,s'}^{(\zeta)}=
 \sum_{j=1}^2\sum_{i=1}^{n_\downarrow}
\sum_{i'=1}^{n_\uparrow}
\frac{ (W^{(j,0)}-\widetilde{W}^{(j,1)(\zeta)})_{2i+s,2i'+s'}}{8\sqrt{n_\uparrow n_\downarrow}}
\label{eta}
\end{equation}
\begin{equation}
\widetilde{W}^{(j,1)(\zeta)}=
e^{i \zeta \frac{\pi}{3}}W^{(j,1)}
+
e^{-i \zeta \frac{\pi}{3}}W^{(j,-1)}.
\label{wideW}
\end{equation}
In Eq. (\ref{eta}), sublattice indexes (A,B) and (a,b) are
 translated to integers $(-1,0)$ in the same way as Fig. 2.

As $E^{[0]}_{\sigma,+}=E^{[0]}_{\sigma,-}$, we perform
 the perturbation calculation for the doubly degenerate states \cite{JJ-Sakurai}.
 The conditions  $\;^t\left(\vec{b}_{\sigma,\tau}^{\;[0](\zeta)}
\right)^*\vec{b}_{\sigma',\tau'}^{\;[0](\zeta)}
=\delta_{\sigma,\sigma'}\delta_{\tau,\tau'}
$ and $V_{(\sigma,+|\sigma,-)}^{(\zeta)}=0$ for this calculation
 require us to choose the factor $f_\sigma^{(\zeta)}$ 
as
\begin{equation}
f_\sigma^{(\zeta)}=
 \frac{\left(w^{(\zeta)}_{\sigma,\sigma}\right)^*}{\left|
w^{(\zeta)}_{\sigma,\sigma}\right|}.
\label{deff}
\end{equation}
The first order formulas are 
\begin{eqnarray}
E^{[1]}_{\sigma,\tau} &=& V_{(\sigma,\tau|\sigma,\tau)}^{(\zeta)}\nonumber \\
 & = & 2\tau \left|w_{\sigma,\sigma}\right| 
\label{energy-1}
\end{eqnarray}
and
\begin{equation}
\vec{b}^{\;[1](\zeta)}_{\sigma,\tau}=\sum_{\tau'=\pm}
\frac{V_{(-\sigma,\tau'|\sigma,\tau)}^{(\zeta)}}{
2E^{[0]}_{\sigma,\tau}}
\vec{b}_{-\sigma,\tau'}^{\;[0](\zeta)}
\label{perturb-wave}
\end{equation}
where we use relation $E^{[0]}_{\sigma,\tau}-E^{[0]}_{-\sigma,\tau'}=2E^{[0]}_{\sigma,\tau}$.
In Eq.(\ref{energy-1}), index $\zeta$ is omitted  as 
$|w_{\sigma,\sigma}^{(+)}|=|w_{\sigma,\sigma}^{(-)}|$.
Using Eqs. (\ref{El0}), (\ref{energy-1}) and $E=E_{\sigma,\tau}^{[0]}+E_{\sigma,\tau}^{[1]}$,
 the wave number $k$  
 is approximated by
\begin{equation}
k_{\sigma,\tau} =  \zeta \frac{ 2}{a}
\left(\frac{\pi}{3}-\sigma \frac{E-2\tau |w_{\sigma,\sigma}|}{\sqrt{3}t} \right)
\label{linear-k}
\end{equation}
 with the  group velocity $\frac{dE}{\hbar dk}= -\zeta\sigma\frac{\sqrt{3}ta}{2\hbar}  $.
The set $\{\vec{b}_{\sigma,\tau}^{(\zeta)} \}$ 
 has a common wave number $k \simeq \zeta 2\pi/(3a)$   while
 we have to prepare   the set $\{ \vec{u}_1, \vec{u}_2, \vec{u}_3, \vec{u}_4 \}$ 
of Eq. (\ref{f-expand}) with  a common energy $E$ and  positive velocities.
Replacing $(E^{[0]}_{\sigma,\tau}, \zeta)$ by $(E, -\sigma)$  in Eq. (\ref{perturb-wave}), we obtain the latter set. The error caused by this replacement is a higher order term and negligible.

Equation (\ref{def-g}) is the repetition of the reduced vector 
$\vec{g}^{\;\prime}_{\xi,\sigma}
\equiv \frac{1}{\sqrt{8n_\xi}}(1,\sigma)$ as $\vec{g}_{\xi,\sigma}=
(\vec{g}^{\;\prime}_{\xi,\sigma},\vec{g}^{\;\prime}_{\xi,\sigma},\cdots,
\vec{g}^{\;\prime}_{\xi,\sigma})$.
Replacing $\vec{g}_{\sigma,\tau}$ by $\vec{g}^{\;\prime}_{\xi,\sigma}$ 
in Eqs. (\ref{e-d}), (\ref{def-dzero}) and (\ref{perturb-wave}),
 $\vec{d}_{\sigma,\tau}^{\;[n](\zeta)} $ is reduced to  the vector  $\vec{d}_{\sigma,\tau}^{\;\prime [n](\zeta) }$. Since we neglect  the evanecent modes, we can use
the simple formula  $^t\vec{c}^{\;(\rm D')}_j=(\;^t\vec{c}_j^{\;\prime (\rm D)} ,\;^t\vec{c}_j^{\;\prime (\rm D)},\cdots,\;^t\vec{c}_j^{\;\prime (\rm D)})$
where 
\begin{equation}
\vec{c}_j^{\;\prime (\rm D)} 
=
\sum_{n=0}^1\Xi^{j+1}
U^{[n]}_{\rm D} 
\vec{\gamma}_+^{\;({\rm D'})} 
+\left( \Xi^{j+1}U^{[n]}_{\rm D}  \right)^*
\vec{\gamma}_-^{\;({\rm D'})} 
\label{cj-D2}
\end{equation}
\begin{equation}
U^{[n]}_{\rm D } =
\left(
\vec{d}_{+,+}^{\;\prime [n] (-)},
\;\vec{d}_{-,+}^{\;\prime [n] (+)},
\;\vec{d}_{+,-}^{\;\prime [n] (-)},
\;\vec{d}_{-,-}^{\;\prime [n] (+)} \right)
\label{cj-Un}
\end{equation}
From Eq. (\ref{linear-k}), we derive
\begin{equation}
 \Xi= \left(\begin{array}{cc}
\Omega^{-1}\Omega_0
, & 0\\
0, &\Omega\Omega_0
\end{array}
\right),\;
\Xi_0=
 \left(\begin{array}{cc}
\Omega_0
, & 0\\
0, &\Omega_0
\end{array}
\right)
\label{def-Lambda}
\end{equation}
where 
\begin{equation}
 \Omega= \left(
\begin{array}{cc}
e^{i \theta_+}
 , & 0\\
0, &e^{i \theta_-}
\end{array}
\right),\;
 \Omega_0 =\left(
\begin{array}{cc}
e^{i\varphi_+}
 , & 0\\
0, &e^{i\varphi_-}
\end{array}
\right)
\label{def-Omega}
\end{equation}
\begin{equation}
\varphi_\sigma =\frac{E}{\sqrt{3}t}+\sigma\frac{2\pi}{3},\;\;\;\theta_\sigma =\frac{2|w_{\sigma,\sigma}|}{\sqrt{3}t}.
\label{theta-phi}
\end{equation}
Equations (\ref{linear-k}) and (\ref{theta-phi}) are related as
 $\varphi_\sigma=\pi+(k_{\sigma,+}+k_{\sigma,-})a/4$
 and  $\theta_\sigma=-(k_{\sigma,+}-k_{\sigma,-})a/4$
 for the positive velocity $\zeta\sigma=-1$.
Though $\Xi_0$ does not appear in Eq. (\ref{cj-D2}), it will be 
 referred to later.
In the relation between Eq. (\ref{f-expand})
 and Eq. (\ref{cj-D2}),
 we should note that $\lambda_l^{\;\rm(D)}=\Xi_{l,l}^2 \neq \Xi_{l,l}
$.
The reduced vectors of single layer regions ($\mu=$ L, R) 
are represented by
\begin{equation}
\vec{c}_j^{\;\prime (\mu )}
=
\frac{1}{2\sqrt{n_{\mu}}}
\left(
\begin{array}{cc}
1, & 1\\
1, &-1
\end{array}
\right)
\sum_{s=\pm} 
\Omega_0^{s(j-j'_\mu)}
\vec{\gamma}_s^{\;(\mu')}.
\label{appendix2-cj}
\end{equation}
where $j'_{\rm l}=j_{\rm l}$ and  $j'_{\rm r}=j_{\rm r}+1$.
From Eqs. (\ref{LD}),(\ref{DR}),(\ref{cj-D2}) and (\ref{appendix2-cj}),
 we derive
\begin{equation}
\left(X^{[0]}_\mu+X^{[1]}_\mu\right)
 \vec{y}_{\rm out}^{\;(\mu)}=-\left(X^{[0]*}_\mu+X^{[1]*}_\mu\right)
\vec{y}_{\rm in}^{\;(\mu)}
\label{scatter-diamond}
\end{equation} 
where outgoing   $\vec{y}_{\rm out}^{\;(\mu)}$ 
 and incoming  $\vec{y}_{\rm in}^{\;(\mu)}$  at boundary $\mu$ 
 are defined by
\begin{equation}
\vec{y}_{\;_{\rm in}^{\rm out}}^{\;(\rm L)}
=\left(\Xi^{\pm (j_{\rm l}+1)}\vec{\gamma}_{\pm}^{\;({\rm D'})},\;
\vec{\gamma}_\mp^{\;({\rm L'})} \right) 
\label{y-L}
\end{equation}
\begin{equation}
\vec{y}_{\;_{\rm in}^{\rm out}}^{\;(\rm R)}=\left(\Xi^{\mp (j_{\rm r}+2)}\vec{\gamma}_{\mp}^{\;({\rm D'})}
,\;\vec{\gamma}_{\pm}^{\;({\rm R'})} \right) .
\label{y-R}
\end{equation}
Substituting   $\vec{y}_{\rm out}^{\;(\mu)}$ in Eq. (\ref{scatter-diamond}) by $\vec{y}_{\rm out}^{\;(\mu)}=(S_\mu^{[0]}+S_\mu^{[1]})\vec{y}_{\rm in}^{\;(\mu)}$,
  we derive 
\begin{eqnarray}
S_\mu^{[0]}
&=& -\left(X_\mu^{[0]}\right)^{-1}
X_\mu^{[0]*},\nonumber \\
S_\mu^{[1]}
&=& -\left(X_\mu^{[0]}\right)^{-1}
(X_\mu^{[1]}S_\mu^{[0]}
+X_\mu^{[1]*}).
\label{XS-0}
\end{eqnarray}
Equation (\ref{XS-0}) enables us to obtain
\begin{eqnarray}
S_{\rm L}&=& 
\frac{1}{2}
\left(
\begin{array}{ccc}
-F^{-2}, & F^{-2}, & \sqrt{2} \bf{1}_2 \\
 F^{-2}, & -F^{-2}, & \sqrt{2} \bf{1}_2 \\
 \sqrt{2} \bf {1}_2, & \sqrt{2} \bf{1}_2, & 0 \\
\end{array}
\right)
\nonumber \\
&& +
\frac{1}{\sqrt{2}E}
\left(
\begin{array}{ccc}
-\alpha_+\sigma_x, & -i \alpha_-\sigma_y, & -F^*G^*\sigma_x
\\
i\alpha_-\sigma_y, & \alpha_+\sigma_x, & F^*G^*\sigma_x\\
 -\sigma_xF^*G^*,
 &  \sigma_xF^*G^*, & 0
\end{array}
\right)
\label{r-0}
\end{eqnarray}
\begin{eqnarray}
S_{\rm R}&=& 
\frac{-1}{2}
\left(
\begin{array}{ccc}
 \bf{1}_2, & \bf{1}_2, &-\sqrt{2}F \\
 \bf{1}_2, & \bf{1}_2, &\sqrt{2}F \\
 -\sqrt{2}F ,& \sqrt{2}F ,& 0 \\
\end{array}
\right)
\nonumber \\
&& +
\frac{1}{\sqrt{2}E}
\left(
\begin{array}{ccc}
-\alpha_+^*\sigma_x, & -i \alpha_-^*\sigma_y, & -\sigma_xG
\\
i\alpha_-^*\sigma_y, & \alpha_+^*\sigma_x, &  -\sigma_x G\\
-G\sigma_x
 &  -G\sigma_x, & 0
\end{array}
\right)
\label{r-1}
\end{eqnarray}
with  the $2 \times 2$ unit matrix  ${\bf 1}_2$,  diagonal matrices
\begin{eqnarray}
F &=& 
\left(
\begin{array}{cc}
f_+^{(-)} & 0 \\
0,  & f_-^{(+)} \\
\end{array}
\right)
\nonumber \\
&= &
\left(
\begin{array}{cc}
e^{iA_+} & 0 \\
0,  & e^{iA_-} \\
\end{array}
\right)
\label{def-F}
\end{eqnarray}
\begin{eqnarray}
G &= &
\left(
\begin{array}{cc}
w_{+,-}^{(+)} & 0 \\
0,  & w_{-,+}^{(-)} \\
\end{array}
\right)
\nonumber \\
&= &
\left(
\begin{array}{cc}
|w_{+,-}|e^{iB_+} & 0 \\
0,  & |w_{-,+}|e^{-iB_-} \\
\end{array}
\right)
\label{def-G}
\end{eqnarray}
Pauli matrices
\begin{equation}
\left \{ \sigma_x,\sigma_y,\sigma_z \right \}=
\left\{
 \left(
\begin{array}{cc}
0, &  1\\
1, & 0
\end{array}
\right),
 \left(
\begin{array}{cc}
0, &  -i\\
i, & 0
\end{array}
\right),
 \left(
\begin{array}{cc}
1, &  0\\
0, & -1
\end{array}
\right)
\right\}.
\label{pauli}
\end{equation}
and
\begin{equation}
\alpha_{\pm}=\frac{1}{\sqrt{2}}
(f_+^{(+)}w_{+,-}^{(+)}\pm f_-^{(-)}w_{-,+}^{(-)} ).
\label{def-xpm}
\end{equation}
 In Eqs. (\ref{def-F}) and (\ref{def-G}), the phases of $f_\sigma^{(-\sigma)}$ and 
$w_{-\sigma,\sigma}^{(+)}$ are 
  denoted by $A_\sigma$ and $B_{-\sigma}$, respectively.
As $(w_{\sigma',\sigma}^{(-)})^*
= w_{\sigma',\sigma}^{(+)}$,
 we omit the index $\zeta$ in the absolute value 
  $|w_{\sigma',\sigma}^{(\zeta)}|$.
See Appendix C for the detail of the calculation.

 In order to combine $S_{\rm L}$ and $S_{\rm R}$ into 
 the $S_{\rm RL}$ matrix of Eq. (\ref{smatrix}),  we partition  
 Eqs. (\ref{r-0}) and (\ref{r-1}) 
 into reflection blocks
 and transmission blocks as
\begin{equation}
S_\mu=
\left(
\begin{array}{cc}
r_\mu^{[0]}
, &
^tt_\mu^{[0]} \\
t_\mu^{[0]}, &
0
 \\
\end{array}
\right)
+
\left(
\begin{array}{cc}
r_\mu^{[1]}
, &
^tt_\mu^{[1]} \\
t_\mu^{[1]}, &
0
 \\
\end{array}
\right)
\label{S-block}
\end{equation}
The transmission matrix $t_{\rm RL}$ 
 in Eq. (\ref{smatrix})
 is represented by
 the superposition  of the multiple reflection waves as
\begin{equation}
t_{\rm RL}= t_{\rm R} 
\Xi^N
\sum_{m=0}^{\infty}\;
\left(r_{\rm L}\Xi^N r_{\rm R}
\Xi^N\right)^m\;^t t_{\rm L} 
\label{multiple-scatter3} 
\end{equation}
 with the overlap length integer $N =j_{\rm r}-j_{\rm l}+1$. 
 The integer $m$ in Eq. (\ref{multiple-scatter3}) 
 is the number of times of the round-trip between $j=j_{\rm l}$  and 
$j=j_{\rm r}$ before the transmission.
Replacing $r_\mu,t_\mu$ by $r_\mu^{[0]},t_\mu^{[0]}$ 
 in Eq. (\ref{multiple-scatter3}),
 we obtain  the zero-order   $t^{[0]}_{\rm RL}$.
 That is a diagonal matrix showing  the diagonal transmission rates
\begin{equation}
T_{\sigma,\sigma}=\frac{4\sin^2(N\theta_{\sigma})
\cos^2\left(A_\sigma-N\varphi_\sigma \right)}
{\cos^4(N\theta_{\sigma})+4\sin^2(N\theta_{\sigma})
\cos^2\left(A_\sigma-N\varphi_\sigma\right)}
\label{T++}
\end{equation}
 with  the phases defined by Eqs. (\ref{theta-phi})  and (\ref{def-F}).

On condition that $\Xi^N \simeq \Xi_0^N$, the
 first order term $t_{\rm RL}^{[1]}$ of Eq. (\ref{multiple-scatter3}) approximates to $ p_1^{(0)}+p_0^{(0)}+ p_1^{(1)}+p_0^{(1)}$ 
where
 \begin{equation}
\left(
\begin{array}{c}
p_n^{(0)} \\
p_n^{(1)}
\end{array}
\right)
=
\left(
\begin{array}{c}
t_{\rm R}^{[n]}\Xi_0^N\;^tt_{\rm L}^{[1-n]}\\
t_{\rm R}^{[0]}\Xi_0^Nr_{\rm L}^{[n]}\Xi_0^Nr_{\rm R}^{[1-n]}
\Xi_0^N\;^tt_{\rm L}^{[0]}
\end{array}
\right).
\label{multiple-scatter2} 
\end{equation}
The superscript $(m)$ and subscript $n$ of $p^{(m)}_n$
 indicate 
the times of the round trip
 and 
the position of the first order matrix, respectively.
The condition $\Xi^N \simeq \Xi_0^N$ 
 is satisfied in the region $N < $ min$(1/|\theta_+|,1/|\theta_-|)=\sqrt{3}t/(2\overline{w})$  where $\overline{w} \equiv $ max$(|w_{+,+}|,|w_{-,-}|)$.
The diagonal elements of Eq. (\ref{multiple-scatter2}) equal zero 
 while the off-diagonal elements of Eq. (\ref{multiple-scatter2}) are
 represented by
\begin{eqnarray}
\left(p_1^{(0)}
\right)_{-\sigma,\sigma}
 & = &
\frac{-w_{-\sigma,\sigma}^{(-\sigma)}}{E}
e^{iN\varphi_\sigma} 
\nonumber \\
\left(p_0^{(0)}
\right)_{-\sigma,\sigma}
 & = &
\frac{-w_{-\sigma,\sigma}^{(\sigma)}}{E}
e^{iN\varphi_{-\sigma}}
\label{A0B0A1B1}
\end{eqnarray}
and $p_n^{(1)}=-\exp\left(i\frac{2NE}{\sqrt{3}t}\right)p_n^{(0)}$. 
From the first order $t_{\rm RL}^{[1]}$, we can 
 derive the off-diagonal transmission rate
\begin{equation}
T_{-\sigma,\sigma}=
16\frac{|w_{-\sigma,\sigma}|^2}{E^2}
\cos^2\left(B_{-\sigma}+\frac{N\pi}{3}\right)
\sin^2\left(\frac{NE}{\sqrt{3}t}\right)
\label{T+-}
\end{equation}
 with the phase $B_{-\sigma}$ defined by  Eq.(\ref{def-G}).
In Eq. (\ref{T+-}),
$-\sigma$ and $\sigma$ correspond to tubes $\uparrow$ (R)
 and $\downarrow$ (L), respectively.

\section{results and discussions}
Firstly we consider the case where $\Delta z =0$ and 
$A_\sigma=B_\sigma =0$.
Figures 5 and 6 show the transmission rates $T_{\sigma',\sigma}$ 
for  the side contact ($E=0.08$ eV)  and the coaxial contact
 ($E=0.3$  eV), respectively, in the case where $n_\downarrow =10$
 and $n_\uparrow =15$. 
 The horizontal axix is  the integer $N=j_{\rm r}-j_{\rm l}+1$.
 The geometrical overlapped length equals  $(N-2)a/2$
 as is shown by Fig. 1.
 Equations (\ref{T++}) and (\ref{T+-}) do not depend on $j_{\rm l}$ 
 when $N$ is fixed. As the author has confirmed that this insensitivity
 to $j_{\rm l}$  also approximately holds in the exact results,
 displayed exact results are limited to the case where $j_{\rm l}=-1$.
The interval of $N$ in each line is three
and the attached numbers 0, 1 and 2 are
 mod($N,3$).
Symbols $(\sigma',\sigma)$ in Fig. 5
 indicate subscripts of $T_{\sigma',\sigma}$.
For the coaxial contact of Fig. 6, $w_{-,\sigma}=0$ and
the exact numerical values of $T_{-,\sigma}$  are negligibly small
 compared to $T_{+,\sigma}$. Thus $T_{-,\sigma}$ is not shown
 in Fig. 6 \cite{footnote}.
In Figs. 5, 6 and other following figures,
the dashed lines represent the approximate formulas
 (\ref{T++}) and
 (\ref{T+-}) while 
 the exact data are shown by solid lines.

The values of Eq. (\ref{overline-w})
 for Figs. 5 and 6 are listed in Table I.
In order to understand  a large difference between
 the side and  coaxial contacts in Table I, 
 we should note  cancellation between $W^{(j,0)}$ and $W^{(j,1)}$   in 
Eq. (\ref{eta}) where $W^{(j,-1)}=W^{(j,1)}$ and $\widetilde{W}^{(j,1)(\pm)}=
W^{(j,1)}$.
This cancellation originates from phase $\pi$ in Eq. (\ref{e-d}).
For reference,  Fig. 7 shows
the interlayer configurations  of the bilayer GR 
 of which the lower 'AB'  and upper 'ab' sublattices are numbered 
 along the armchair chain.
 In  Fig. 7(a), 
 A1-a1, B1-b1 and B1-a3 elements of   $W^{(j,0)}$ 
 cancel  A1-a2, B1-b2 and B1-a2 elements of   $W^{(j,1)}$ 
 completely.
 Thus only the A1-b1 element of $W^{(j,0)}$ contributes to Eq. (\ref{overline-w}) and $w_{\sigma',\sigma}= \sigma' \eta_{\rm Ab}$.
 It indicates that only the vertical  bonds  contribute
 to Eq. (\ref{overline-w}).
 In the same way,  $w_{\sigma',\sigma}= \sigma \eta_{\rm Ba}$
 in Fig. 7(b)
 and
 $w_{\sigma',\sigma}= (1+\sigma\sigma') \eta_{\rm Aa}$
 in  Fig. 7(c).
  Since  Fig. 2 is
 similar to Fig. 7 in the local configuration,
 vertical bonds indicated by ovals are dominant in Eq. (\ref{overline-w})
 where all the vertical bonds have similar positive values
 in $W^{(j,0)}$.
As is shown in Fig. 2,   the number of the vertical bonds
 in Eq. (\ref{overline-w})
 is considerably larger in the coaxial contact 
 than in the side contact.
 This is the reason why
$w_{+,+}$  of the coaxial contact 
 is  remarkably larger than $w_{+,+}$ of the side contact.
In the side contact, the interlayer bonds
 are limited to  the contact line  $\theta^\uparrow \simeq \theta^\downarrow \simeq 0$ with the Ab stacking,  namely, 
 $w_{\sigma',\sigma} \simeq \sigma'\eta_{\rm A,b}$.
 In the rest of this paragraph, we discuss  the coaxial contact.
 In contrast to the side contact, the vertical bonds apprear in all
 the four terms in Eq. (\ref{overline-w}).
 As  the vertical bonds have similar lengths, 
 the four $\eta$'s are  close to each other.
 It explains the relation  $w_{+,+} > |w_{+,-}|,|w_{-,+}|,|w_{-,-}|$.
Relations $(\eta_{\rm A,a},\eta_{\rm A,b})=(\eta_{\rm B,a},\eta_{\rm B,b})$ 
 and $w_{+,-}=w_{-,-}=0$ hold on condition that mod($n_\uparrow$,3) =0 and 
$|n_\downarrow-n_\uparrow|=5$.
  This vanishing of $w$ is called
 the threefold cancellation in Ref. \cite{sawai}.
 In Fig. 2(b), for example, $\Box$, $\Diamond$ and $\triangle$
 bonds cancel  $\Box'$, $\Diamond'$ and $\triangle'$, respectively.
 Whether the three fold cancellation occurs or not, 
 $w_{+,+}$ is dominant among the four $w$'s.
Here we should remember that Eq. (\ref{T+-}) has been derived
 under the condition $N < \sqrt{3}t/(2\overline{w})$. 
The difference between the two contacts in $\overline{w}=$
 max($|w_{+,+}|,|w_{-,-}|)$ appears in maximum $N$ for the
 effectiveness of Eq. (\ref{T+-}). 
Namely, coincidence between
 solid and dashed lines is limited to region $N < 20$
 in Fig. 6(b), while
 that is seen in the wider range $N < 100$
 in Fig. 5(b).
 Considering that Eq. (\ref{T++}) reaches
 unity at $N=\sqrt{3}t\pi/(4w_{\sigma,\sigma})$,
 we notice that
  approach of  Eq. (\ref{T++}) to unity loses 
  effectiveness of Eq. (\ref{T+-}).
 On the other hand,
  effectiveness of Eq. (\ref{T++})
 is not influenced by Eq. (\ref{T+-})
 as is shown in Fig. 5(a) and Fig. 6(a).
With a fixed $N$, Eq. (\ref{T+-}) reaches its maximum 
$16\cos^2(N\pi/3)w_{-\sigma,\sigma}^2N^2/(3t^2)$ 
at $E=0$. Thus the maximum of Eq. (\ref{T+-})
 in its effective range  $N < \sqrt{3}t/(2\overline{w})$  is estimated to be 
$4w_{-\sigma,\sigma}^2/\overline{w}^2$.
As $w_{-\sigma,\sigma}^2/\overline{w}^2$ is remarkably larger
 in the side contact
 than in the coaxial contact,
 we concentrate our attention on the side contact below.

Dependence of Eq. (\ref{T++}) on $N$ is determined by the phases
 $N\theta_\sigma$ and $N\varphi_\sigma$.
As a function of $N$,
the former and the latter correspond to slow and
 rapid oscillations, respectively.
Connecting data points with the interval of three, the rapid oscillation
 is smoothed in Fig. 5.
Since $\theta_\sigma$ is independent of $E$, only $\varphi_\sigma$ 
 determines the dependence of Eq. (\ref{T++}) on $E$.
In Fig. 5(a), the line $(\sigma,\sigma)$-1 is similar to
 the line $(-\sigma,-\sigma)$-2 in the period
 since mod($N\varphi_\sigma,2\pi)= \frac{2\pi}{3}\sigma$mod$(N,3)+\frac{EN}{\sqrt{3}t}$. 
The first nodes of  $(\sigma,\sigma)$-0 in Fig. 5(a) 
and the first peaks in Fig. 5(b) have the common horizontal position 
$N=\pi\sqrt{3}t/(2|E|) \simeq 93$.

 Figure 8  shows (a) $\overline{T}_{+,-}$
 and (b) Landauer's formula conductance 
 $\sum_{\sigma',\sigma} \overline{T}_{\sigma',\sigma}$
 for the energies $E=0.05, 0.08, 0.1, 0.15$ eV
 where $\overline{T}(N) \equiv \frac{1}{3}\sum_{j=-1}^1T(N+j)$
  denotes the 'smoothed'  transmission rate. 
In the transformation of $T$ into $\overline{T}$,  the rapid oscillation 
 with the wave length $3a/2$ is smoothed out.
 Effectiveness of  Eq. (\ref{T+-})  is confirmed for the energies
 $E=0.15, 0.1, 0.08$ eV in Fig. 8(a).
 The peak positions of solid lines are consistent with
 those of dashed lines $(N,\overline{T}_{+,-})=(\pi\sqrt{3} t/(2|E|),8w_{+,-}^2/E^2)$.
 As will be clarified latter, this peak is important
 for the smoothed Landauer's formula conductance
 in Fig. 8(b).
When $E=$ 0.05 eV, however, the solid lines are suppressed compared to
 the dashed line in Fig. 8.  
 This suppression is also found in
 Fig. 9 showing
 $T_{\sigma',\sigma}$  as a function of $E$ 
 with $N= 81, 82$.
 In Fig. 9,  the approximate formulas satisfactorily reproduce the exact
 results  except overestimation of
 the peak height at $(N,E)=(81,0)$.
 This suppression of the zero energy peak 
 is caused by the pseudogap.
 As Eq. (\ref{linear-k})
 shows no gap, 
$\overline{N}_{\rm D} =4$ in the perturbation calculation.
 On the other hand, pseudogap regions $\overline{N}_{\rm D} =2$ 
 appear near $E=0$ in the exact dispersion lines 
 as is shown by Fig. 10. Compared to the pseudogap,
 the width of the real gap $\overline{N}_{\rm D} =0$ is negligibly small.
 The solid lines are similar to
 the dashed lines in the energy difference  between the neighboring lines
 while  crossing occurs only in the dashed lines.
 Thus the pseudogap width  is estimated to be $4\overline{w}$. 
 Since  Eq. (\ref{T+-})
 is effective outside the pseudogap  $|E| > 4\overline{w}$,
  the maximum of Eq. (\ref{T+-}) is estimated to be
  $w_{-\sigma,\sigma}^2/\overline{w}^2$.
 Outside the pseudogap, Eq. (\ref{T+-}) can reach its maximum
 at $N=\pi\sqrt{3}t/(2|E|)$ in its effective range $N < \sqrt{3} t/(2\overline{w})$.
 The diagonal $T_{\sigma,\sigma}$ 
 has zero energy peak
 only when mod($N,3$)=0, 
 while off-diagonal  $T_{-\sigma,\sigma}$ has it irrespective of
 mod($N,3$). This difference between $T_{\sigma,\sigma}$ 
 and $T_{-\sigma,\sigma}$  becomes more obvious in Fig. 11
 showing  the smoothed $\overline{T}$ with $N=82$ 
  as a function of $E$.
 The zero energy peaks of $\overline{T}_{\sigma,\sigma}$
 are replaced by the dips 
 while those of $\overline{T}_{-\sigma,\sigma}$ resist the suppression by the pseudogap.
 We can also find that the rise of the conductance with lowered $E$ in Fig. 8(b)   comes from the off-diagonal part  $T_{+,-}+T_{-,+}$,
 although  $T_{+,-}+T_{-,+}$ is less than the diagonal part $T_{+,+}+T_{-,-}$ 
 in  Fig. 11 outside the pseudogap.

The analytical formulas (\ref{T++}) and (\ref{T+-})
 enable us  to  discuss the $|W|^2$ and $(k_1,k_2)$ characteristics 
 mentioned in  Sec. I.
When $\Delta z=0$, $N \ll \sqrt{3}t/|w_{\sigma,\sigma}|$
 and $N \ll \sqrt{3}t/|E|$, Eqs. (\ref{T++}) and (\ref{T+-}) 
 are unified into  $\frac{16}{3}(w_{\sigma',\sigma}/t)^2N^2
\cos^2(N\pi/3)$. 
It clearly indicates that all four parameters $w_{\sigma',\sigma}$
 have the same $|W|^2$ characteristic.
As a function of the overlapped length $Na/2$, 
Eqs. (\ref{T++}) and (\ref{T+-}) show superposition of the rapid and slow oscillations. 
It can be
 considered as a beat with the wave number Eq. (\ref{linear-k}).
The periods of Eq. (\ref{T++}) are consistent with $|k_{\sigma,+}-k_{\sigma,-}|=4\theta_\sigma/a$ and $|k_{\sigma,+}+k_{\sigma,-}|=4|\varphi_\sigma-\pi|/a$.
 In the same discussion on the off-diagonal transmission,
 however, we are not clear how to choose $(\tau,\tau')$ 
 in the calculation of  $|k_{+,\tau}-k_{-,\tau'}|$
 and $|k_{+,\tau}+k_{-,\tau'}|$. 
 Neglecting $w_{\sigma,\sigma}$ in Eq. (\ref{linear-k}), we can obtain 
 approximations  $|k^{(-)}_{+,\tau}-k_{-,\tau'}^{(+)}| \simeq 4\pi/(3a)$
 and $|k^{(-)}_{+,\tau}+k_{-,\tau'}^{(+)}| \simeq 4|E|/(\sqrt{3}ta)$
 that agree with the periods of Eq. (\ref{T+-}).
 Here we explicitly show the index $\zeta$ in superscripts
 of $k_{\sigma,\tau}$  for the explanation.

Figure  12 illustrates
the multiple reflection between the two boundaries $j_{\rm l}$ and $j_{\rm r}$ 
 with the notation
 of Eq. (\ref{multiple-scatter2})  in the case where symmetric (+) channel  is incident from region L.
 The circles and triangles  represent
  transmission $t_\mu$ and the reflection $r_\mu$ 
 at $j=j_\mu$, while the closed and open symbols
 correspond to the first and zeroth order, respectively.
 The rectangles indicate
 the  phase $N \varphi_\sigma $ accumulated in $\sigma$ channel
 along a one-way path
 either $j_{\rm l} \rightarrow j_{\rm r}$ 
 or $j_{\rm l} \leftarrow j_{\rm r}$.
  The $+$  channel (dashed line path)  changes  into the $-$ 
 channel (solid line path)
  after an encounter with the closed symbol.
 Relative phases between $p_1^{(m)}$
  and $p_0^{(m)}$ with a common $m$ are 
 $(N\varphi_+ -B_-)$  and $(N\varphi_- + B_-)$ 
 where the phase $B_-$ comes
 from the closed symbols.
  It explains the factor 
 $|e^{i(N\varphi_+ -B_-)} +e^{i(N\varphi_- + B_-)}|^2 =4\cos^2(B_-+ \pi N /3)$
 in Eq.(\ref{T+-}).
 Compared to the $p_m^{(0)}$ path, on the other hand, the $p_m^{(1)}$ path has an additional round trip with the phase  factor $e^{i(\varphi_-+\varphi_+)N}$.
 At the same time, we also have to consider factor $(-1)$
 in the relations 
 $ t_{\rm R}^{[0]}r_{\rm L}^{[1]}r_{\rm R}^{[0]} =- t_{\rm R}^{[1]}$ 
 and 
$ r_{\rm L}^{[0]}r_{\rm R}^{[1]} \!^tt_{\rm L}^{[0]} =-^tt_{\rm L}^{[1]}$.
With these factors, we see the factor
$|1-e^{i(\varphi_-+\varphi_+)N}|^2=4\sin^2( NE /\sqrt{3} t)$ 
  in Eq.(\ref{T+-}).
The analytical formulas (\ref{T++}) and (\ref{T+-})
 are effective for  general $\Delta z$ and $\Delta \theta$.
Figures  13 and 14 show the transmission rate $T_{\sigma',\sigma}$ 
 as a function of $\Delta \theta$  and 
$\Delta z$, respectively, in the case where $(n_\downarrow,n_\uparrow) =(10,15), N=82 , E=$ 0.05 eV. In Figs. 13 and 14, $\Delta z$ and $\Delta \theta$ are fixed to 
 zero, respectively.
 In Fig. 13, 
 the off-diagonal transmission rate vanishes
 at  $\Delta\theta = -\pi/(3n_\uparrow),2\pi/(3n_\uparrow)$
 with  the common mirror plane.
The exact results are reproduced well by Eqs. (\ref{T++}) and (\ref{T+-})
 also  for the dependence on $\Delta \theta$ and $\Delta z$.
Although the phases $A_\sigma$ and $B_\sigma$ are irrelevant
 to the band structure (\ref{linear-k}),
 they are essential for the dependence of  Eqs. (\ref{T++}) and (\ref{T+-})
 on  $\Delta z$.
The data are shown 
for the discrete values
 $\Delta \theta =m\pi/(150n_\uparrow) $ 
and  
$\Delta z =ma/400$ 
 with integers $m$.
The discontinuous change in Figs. 13 and  14 comes
 from the cut-off radius $r_c$ of the interlayer Hamiltonian $W$.
If more realistic interlayer Hamiltonian were used,
 the lines would be continuous.
We choose the range $|\Delta z| < 0.015$ nm in Fig. 14
 because we have to consider $W^{(j,\pm 2)}$ outside the range.

When $N > \sqrt{3}t/(2\overline{w})$, 
  the approximation $\Xi^N \simeq \Xi_0^N$  becomes invalid
 and   many  terms other than  Eq. (\ref{multiple-scatter2}) 
  contribute to $t_{\rm RL}^{[1]}$.
  It is the reason why
 random oscillation replaces
 Eq. (\ref{T+-})  when  $N > \sqrt{3}t/(2\overline{w})$.
 It  corresponds to the case where we cannot neglect 
 ambiguity about  $(\tau,\tau')$  in the discussion on the $(k_1,k_2)$ characteristic.
The $(k_1,k_2)$  characteristic appears in both Eqs. (\ref{T++}) and (\ref{T+-}) in this way,
 but the absolute values of the off-diagonal parameters $|w_{+,-}|,|w_{-,+}|$
  are   irrelevant to it.
On the other hand,
  we cannot derive the maximum of  transmission rate
 from  the $(k_1,k_2)$ characteristic.
 The effect of Eq. (\ref{energy-1}) on $S_\mu^{[1]}$  
 can be neglected as higher order 
 when $|E| (\simeq \left|E^{[0]}\right|)$
 is much larger than $ \left|E^{[1]}\right|$.
This condition $ \left|E^{[0]}\right| \gg \left|E^{[1]}\right|$
 corresponds to the outside of the pseudogap $|E| > 4\overline{w}$.
Accordingly only the off-diagonal parameters  $w_{+,-}$  and $w_{-,+}$
 appear in Eq. (\ref{T+-})
 while  they have no relation to Eq. (\ref{linear-k}).
 Conversely  the diagonal $\omega_{\sigma,\sigma}$ 
 is irrelevant to Eq. (\ref{T+-}), though
 it determines the energy shift
 (\ref{energy-1}) and the dispersion (\ref{linear-k}). 
 As $\omega_{+,-}$ and  $\omega_{-,+}$ cannot be detected by
 the energy spectrum,
 the measurement of the off-diagonal transmission rate (\ref{T+-})
 will enrich  our understanding of the interlayer Hamiltonian.

Formulas similar to Eq. (\ref{T++})
 have been reported in Refs. \cite{10-side-contact-ref-1}  and \cite{ref-2}.
The parameters $k, \kappa $ and $L$  of Ref. \cite{10-side-contact-ref-1} are related to those of  Eq. (\ref{T++}) as
 $k=2\varphi_\sigma/a, \kappa =2\theta_\sigma/a, L=Na/2$.
Replacing $\epsilon, \cos(k_1-k_2)L $ and $\sin\left[(k_1+k_2)\frac{L}{2}+\theta \right]$
 by $1/2, \sin\left[(k_{\sigma,+}-k_{\sigma,-})\frac{Na}{4}\right] $
 and $\cos\left[A_\sigma-(k_{\sigma,+}+k_{\sigma,-})\frac{Na}{4} \right] $,
 respectively, we can transform the formula of Ref. \cite{ref-2} into Eq. (\ref{T++})  .
 The formulas, however, are not explicitly related to the TB Hamiltonian elements and energy in Refs. \cite{10-side-contact-ref-1} and \cite{ref-2}.
 The explicit relation shown by Eqs. (\ref{deff}), 
 (\ref{theta-phi}) and (\ref{def-F}) makes their discussions
  quantitative
 and is also essential in our discussion.
 Furthermore
 we also present the analytical formula of the off-diagonal transmission rate (\ref{T+-}) which has been neglected  so far in other works.
 It is clarified that Eq. (\ref{T+-}) is more significant
  than Eq. (\ref{T++}) for the zero energy peak in the side contact.
 The analytical calculation  for the  zigzag NT junctions is
 complicated  since  the reduction of the vector dimension
 $\vec{g} \rightarrow \vec{g}^{\;\prime},
\vec{b} \rightarrow \vec{d} \rightarrow \vec{d}^{\;\prime},
\vec{c} \rightarrow \vec{c}^{\;\prime}$ in Sec. \ref{sec-appro}
 is impossible.
 This difficulty might be overcome by 
 the effective mass theory
 and is left for a future study.
 Though the TB Hamiltonian is only a first guess,
 Eqs. (\ref{T++}) and (\ref{T+-}) can be applied to more precise one
 derived from the first principle calculation
 with geometrical optimization
  because
 our systematic approximation is 
 free from 'fitting parameters'
 in a sense that $w_{\sigma',\sigma}$ 
 is uniquely determined by the Hamiltonian.

\appendix
\section{symmetry of $S$ matrix and normalization}
 The TB equation is represented by 
\begin{eqnarray}
\;^tQ_1^{(m+1)}\vec{f}_{m+1}+Q_0^{(m)}\vec{f}_l+Q_1^{(m)}\vec{f}_{m-1}
&=& E\vec{f}_{m} \label{app-TB1}
\\
&=& i\hbar \frac{\partial }{\partial t} \vec{f}_m
\label{app-TB2}
\end{eqnarray}
where 
$^t\vec{f}^{\;(\mu)}_{m}
\equiv 
(\;^t\vec{c}^{\;(\mu)}_{2m-1}, \;^t\vec{c}^{\;(\mu)}_{2m}) $.
When  $1 \leq m \leq \frac{N}{2}$,
\begin{equation}
Q_0^{(m)}=
\left(
\begin{array}{cc}
H^{(1,0)}, & 
H^{(1,1)}
\\
^tH^{(1,1)}
, & H^{(2,0)}
\end{array}
\right)
\label{app-H0}
\end{equation}
with $H^{(j,\Delta j)}$ defined by Eq. (\ref{def-H}).
When  $2 \leq m \leq \frac{N}{2}$,
\begin{equation}
Q_1^{(m)}=
\left(
\begin{array}{cc}
0, & H^{(1,1)}
\\
0, & 0
\end{array}
\right).
\label{app-H1}
\end{equation}
Deleting unnecessary blocks from $H^{(j,\Delta j)}$ in Eqs. (\ref{app-H0}) and 
(\ref{app-H1}),  we can obtain $Q_0^{(m)}$ and $Q_1^{(m)}$  for other values of $m$.
Equations (\ref{app-TB1}) and  (\ref{app-TB2}) enable us to derive
the conservation of  the probability $  0=-J_{m+1}+J_{m} $ 
 and $\frac{\partial }{\partial t} |\vec{f}_l|^2=-J_{m+1}+J_{m}
$, respectively,
 with the probability flow
\begin{equation}
J_{m} \equiv \frac{2}{\hbar}{\rm Im}(\;^t\vec{f}_{m}^{\;*}Q_1^{(m)}\vec{f}_{m-1})
\label{app-J-def}
\end{equation}
 between $z=(m-1)a$ and $z=ma$.
As we discuss the steady state corresponding to Eq. (\ref{app-TB1}),  $J_{m}$ does not depend on $m$.

Using Eq. (\ref{f-expand}), we obtain
\begin{equation}
J_{m}=\frac{2}{\hbar}{\rm Im} \left[ \sum_{l,l'} I_{l',l}
(\lambda_{l'}^*\lambda_l
)^{m}\gamma_{l'}^*\gamma_l \right]
\label{app-J}
\end{equation}
where
\begin{equation}
I_{l',l}\equiv \;^t\vec{u}_{l'}^{\;*}
Q_1
\vec{u}_l\lambda_l^{-1} .
\label{app-I-def}
\end{equation}
 Since $\vec{f}_m=\lambda_l^m\vec{u}_l$ is a solution of Eq. (\ref{app-TB1}),
\begin{equation}
(Q_0-E+\lambda_l\;^tQ_1+\lambda_l^{-1}Q_1)
\vec{u}_l=0.
\label{app-I-TB}
\end{equation}
Multiplying   $^t\vec{u}_{l'}^{\;*}$ by Eq. (\ref{app-I-TB}), we derive
\begin{equation}
\;^t\vec{u}_{l'}^{\;*}(Q_0-E)\vec{u}_{l}+\lambda_{l'}^*\lambda_lI_{l,l'}^*+I_{l',l}=0.
\label{app-J-TB2}
\end{equation}
Exchanging $l$ and $l'$ in complex conjugate of  Eq. (\ref{app-J-TB2}),
 we obtain
\begin{equation}
\;^t\vec{u}_{l'}^{\;*}(Q_0-E)\vec{u}_{l}
+\lambda_{l'}^*\lambda_lI_{l',l}+I_{l,l'}^*=0.
\label{app-TB3}
\end{equation}
Eliminating $I_{l,l'}^*$ in Eqs. (\ref{app-J-TB2}) and (\ref{app-TB3}),
we obtain
\begin{equation}
\left[ 1-(\lambda_l\lambda_{l'}^*
)^2\right]I_{l',l}=(\lambda_l\lambda_{l'}^*-1)
\;^t\vec{u}_{l'}^{\;*}(Q_0-E)\vec{u}_{l}.
\label{app-I-2}
\end{equation}
Equation (\ref{app-I-2})  indicates that
$I_{l,l'}=I_{l',l}^*$ except when 
\begin{equation}
\lambda_l\lambda_{l'}^* =1.
\label{app-I-cond}
\end{equation}
Thus  only the terms satisfying Eq. (\ref{app-I-cond}) contribute to Eq. (\ref{app-J}) being independent of $m$.
When $l=1,2,\cdots, \overline{N}_{\mu}$,
 $\vec{u}_l$ is normalized
as 
\begin{equation}
{\rm Im}(I_{l,l})=\pm \frac{\sqrt{3}}{4}t
\label{app-norm}
\end{equation}
where double signs $\pm$ are consistent with those of $l$. 
The constant $J_m$ with the normalization (\ref{app-norm}) is represented by
\begin{eqnarray}
J & =& 
\frac{\sqrt{3}t
}{2\hbar}
\sum_{l=1}^{\overline{N}_{\rm L}}|\gamma_l^{\;\rm (L)}|^2-
|\gamma_{-l}^{({\rm L})}|^2
\label{app-J-L}
\\
&=&
\frac{\sqrt{3}t}{2\hbar}
\sum_{l=1}^{\overline{N}_{\rm R}}|\gamma_l^{\;\rm(R)}|^2-
|\gamma_{-l}^{({\rm R})}|^2
\label{app-J-R}
\\
&=& 
J^{\;\rm (D)}_{\rm  eva}
+\frac{\sqrt{3}t}{2\hbar}
\sum_{l=1}^{\overline{N}_{\rm D}}|\gamma_l^{\;\rm (D)}|^2-
|\gamma_{-l}^{\;\rm (D)}|^2.
 \label{app-J-D}
\end{eqnarray}
In Eq. (\ref{app-J-D}),
\begin{equation}
J^{\;\rm (D)}_{\rm  eva}
\equiv \frac{2}{\hbar}\sum_{l> \overline{N}_{\rm D}}^{2n_{\rm D}}
{\rm Im} \left(
I_{l,l'}^{\rm (D)}\gamma_{l}^{\;\rm (D)}\gamma_{l'}^{\;\rm (D)*}
+I_{l',l}^{\rm (D)}\gamma_{l'}^{\;\rm (D)}\gamma_{l}^{\;\rm (D)*}
\right) 
\label{app-J-D-eva}
\end{equation}
comes from the evanescent modes
where $l'$ is less than $-\overline{N}_{\rm D}$ and determined by Eq. (\ref{app-I-cond}).
Equations (\ref{app-J-L}) and (\ref{app-J-R})  indicate the  relation
$|\vec{\gamma}^{\;(\rm L')}_+|^2+|\vec{\gamma}^{\;(\rm R')}_-|^2=
|\vec{\gamma}^{\;(\rm L')}_-|^2+|\vec{\gamma}^{\;(\rm R')}_+|^2$ 
  that is equivalent to the unitarity $\;^tS_{\rm RL}^*=S^{-1}_{\rm RL}$.

The wave function $\Psi$ is approximated by linear combination
 of real and orthonormal  $\pi$  orbitals $\phi_{j,i}^{(\xi)}$.
When 
$\Psi=\sum_{i,j}\sum_{\xi=\uparrow,\downarrow}
 c_{j,i}^{(\xi)} \phi_{j,i}^{(\xi)}$
  satisfies the  Schroedinger equation,
$\Psi^*=\sum_{i,j}\sum_{\xi=\uparrow,\downarrow}
 c_{j,i}^{(\xi)*} \phi_{j,i}^{(\xi)}$ also does.
It indicates compatibility between
 Eq. (\ref{smatrix}) 
 and
\begin{equation}
\left(
\begin{array}{c}
\vec{\gamma}^{\;(\rm L')*}_+
\\
\vec{\gamma}^{\;(\rm R')*}_-
\end{array}
\right)
=
\left(
\begin{array}{cc}
r_{\rm LL}, & t_{\rm LR} 
\\
t_{\rm RL}, & r_{\rm RR}\\
\end{array}
\right)
\left(
\begin{array}{c}
\vec{\gamma}^{\;(\rm L')*}_-
\\
\vec{\gamma}^{\;(\rm R')*}_+
\end{array}
\right)
\label{app-smatrix2}
\end{equation}
that is equivalent to relation  $S_{\rm RL}^{-1}=S_{\rm RL}^*$.
As $S_{\rm RL}$ is also unitary ($S_{\rm RL}^{-1}=\;^tS_{\rm RL}^*$), $S_{\rm RL}$ is symmetric $(\;^tS_{\rm RL}=S_{\rm RL}$).
In the single junction with the infinite length of region D,
 $J_{\rm eva}^{\rm (D)}=0$ because  either $\gamma_l^{(\rm D)}$ or $\gamma_{l'}^{(\rm D)}$  must be zero  in Eq. (\ref{app-J-D-eva})
 to avoid the divergence in region D.
Since $S_\mu$ corresponds to the single junction with  zero $J_{\rm eva}^{\rm (D)}$,
$S_\mu$  is also  symmetric and unitary in the same way as $S_{\rm RL}$.
However, it should be noted that
 $J_{\rm eva}^{\rm (D)}$ is {\it not} zero for the double junction
 L-D-R with a finite length of region D.
The exact calculation of $S_{\rm RL}$ includes the effect of  Eq. (\ref{app-J-D-eva}) as is explicitly shown by Appendix B.

For the propagating waves $l=\pm 1, \pm 2,\cdots, \pm \overline{N}$,
 we can derive  
\begin{equation}
\;^t\vec{u}_l^{\;*}
H(k)\vec{u}_l = E|\vec{u}_l|^2
\label{app-Hk}
\end{equation}
 from  Eq. (\ref{app-TB1})
 where $\lambda_l=e^{ika}$ and
\begin{equation}
H(k)= (Q_0+
\;^tQ_1e^{ika}+Q_1e^{-ika}).
\label{app-Hk-def}
\end{equation}
In Sec. \ref{sec-appro}, Eq. (\ref{app-Hk-def}) is denoted by $H_0+V$.
Differentiating Eq. (\ref{app-Hk}), we obtain
\begin{equation}
\;^t\vec{u}_l^{\;*} \frac{d H(k)}{dk} \vec{u}_l = \frac{d E}{dk}|\vec{u}_l|^2
\label{app-Hk-diff}
\end{equation}
where we use the relations
$\frac{d\;^t\vec{u}_l^{\;*}}{dk} H(k) \vec{u}_l = E\frac{d\;^t\vec{u}^{\;*}_l}{dk} \vec{u}_l$
and
$\;^t\vec{u}_l^{\;*} H(k)\frac{d\vec{u}_l}{dk} =  E\;^t\vec{u}_l^{\;*} \frac{d\vec{u}_l}{dk}$.
 From  Eqs. (\ref{app-I-def}),  (\ref{app-Hk-def}) and
(\ref{app-Hk-diff}), we derive  
\begin{equation}
2a{\rm Im}(I_{l,l})  = \frac{d E}{dk}|\vec{u}_l|^2.
\label{app-Hk-I}
\end{equation}
Equation (\ref{app-Hk-I}) shows that the probability flow ${\rm Im}(I_{l,l})$
and the group velocity $\frac{dE}{dk}$ have the same sign.
 Normalization 
\begin{equation}
|\vec{u}_l|^2=1
\label{app-norm-2}
\end{equation}
 used in Sec. \ref{sec-appro} 
 is an approximation to normalization (\ref{app-norm})
 where  the group velocity $\frac{d E}{dk}$ is
 approximated as $\pm \frac{\sqrt{3}}{2}ta$.
 In the exact calculation of Sec. \ref{sec-exact}, however, we use Eq. (\ref{app-norm})  while
 Eq. (\ref{app-norm-2}) is {\it not} used.

\section{exact numerical calculation}
The transfer matrix derived from (\ref{recursion}) is represented by
\begin{eqnarray}
\Gamma^{(\mu)}
=
\left(
\begin{array}{cc}
-\spadesuit_2, &
\diamondsuit_2^{(\mu)} \\
-\diamondsuit_1^{(\mu)}\spadesuit_2, 
&
-\spadesuit_1+\diamondsuit_1^{(\mu)} \diamondsuit_2^{(\mu)}
\\
\end{array}
 \right)
\end{eqnarray}
where 
$ h_\mu^{(j,1)}\diamondsuit_j^{(\mu)}= E{\bf 1} -h_\mu^{(j,0)}$
 and
$ h_\mu^{(j,1)}\spadesuit_j^{(\mu)}= h_\mu^{(j,-1)}$
 with the notatin $h_{\rm L}=h_\downarrow,
h_{\rm R}=h_\uparrow$ 
 and $h_{\rm D}=H$.
Though $\spadesuit_j^{(\rm L)}$  and $\spadesuit_j^{(\rm R)}$
 are equivalent to the unit matrices,
$\spadesuit_j^{(\rm D)}  \neq \bf {1}$   when
 $\Delta z  \neq 0$.
When we allocate Eq. (\ref{f-expand}) to $\vec{c}_j$ as
\begin{equation}
\vec{c}_j=\left\{\begin{array}{ll}
 \vec{c}_j^{\;(\rm L)}  & \mbox{ $ (j \leq j_{\rm l})$} \\
 \vec{c}_j^{\;(\rm D)}  & \mbox{ $ (j_{\rm l}+1 \leq j \leq j_{\rm r})$} \\
 \vec{c}_j^{\;(\rm R)}  & \mbox{ $ (j_{\rm r}+1 \leq j)$}  .
\end{array}
\right. 
\label{app-cj}
\end{equation}
  TB equations at the boundaries $j=j_{\rm l},j_{\rm r}$ are represented by
\begin{eqnarray}
E\vec{c}^{\;(\rm L)}_{j_{\rm l}}
&=&
h_\downarrow^{(j_{\rm l} ,1)}
\vec{c}^{\;(\rm L)}_{j_{\rm l}-1}
+h_\downarrow^{(j_{\rm l} ,0)}\vec{c}^{\;(\rm L)}_{j_{\rm l}}+
h_\downarrow^{(j_{\rm l} ,1)}
\vec{c}^{\;(\rm D,\downarrow)}_{j_{\rm l}+1} \nonumber \\
& &+W^{(j_{\rm l} ,1)}\vec{c}^{\;(\rm D,\uparrow)}_{j_{\rm l}+1}
\label{app-1} 
\end{eqnarray}
\begin{eqnarray}
E\vec{c}^{\;(\rm D)}_{j_{\rm l}+1}
&=&
\left(
\begin{array}{c}
h_\downarrow^{(j_{\rm l}+1 ,1)}
\\
\;^tW^{(j_{\rm l} ,1)}
\end{array}
 \right)\vec{c}^{\;(\rm L)}_{j_{\rm l}}
+H^{(j_{\rm l}+1 ,0)}\vec{c}^{\;(\rm D)}_{j_{\rm l}+1}
 \nonumber \\
& &
+H^{(j_{\rm l}+1 ,1)}\vec{c}^{\;(\rm D)}_{j_{\rm l}+2}
\label{app-2} 
\end{eqnarray}

\begin{eqnarray}
E\vec{c}^{\;(\rm D)}_{j_{\rm r}}
&=&
H^{(j_{\rm r} ,-1)}\vec{c}^{\;(\rm D)}_{j_{\rm r}-1}
+H^{(j_{\rm r},0)}\vec{c}^{\;(\rm D)}_{j_{\rm r}}
\nonumber \\
& &
+\left(
\begin{array}{c}
W^{(j_{\rm r} ,1)} \\
\\
h_\uparrow^{(j_{\rm r} ,1)}
\end{array}
 \right)\vec{c}^{\;(\rm R)}_{j_{\rm r}+1}
\label{app-3} 
\end{eqnarray}

\begin{eqnarray}
E\vec{c}^{\;(\rm R)}_{j_{\rm r}+1}
&=&
\;^tW^{(j_{\rm r} ,1)}\vec{c}^{\;(\rm D,\downarrow)}_{j_{\rm r}}
+h_\uparrow^{(j_{\rm r}+1 ,1)}\vec{c}^{\;(\rm D,\uparrow)}_{j_{\rm r}}
\nonumber \\
& &
+
h_\uparrow^{(j_{\rm r}+1 ,0)}\vec{c}^{\;(\rm R)}_{j_{\rm r}+1}
+h_\uparrow^{(j_{\rm r}+1 ,1)}\vec{c}^{\;(\rm R)}_{j_{\rm r}+2}.
\label{app-4} 
\end{eqnarray}
Since $\vec{c}_j^{\;(\mu)}$ of Eq. (\ref{f-expand}) satisfies
 Eq. (\ref{recursion})
and
\begin{equation}
E
\left(\begin{array}{c}
\vec{c}_j^{\;\rm (L)} \\
\vec{c}_j^{\;\rm (R)}
\end{array}
\right)
=\sum_{\Delta j=-1}^1 
\left(
\begin{array}{c}
h^{(j,\Delta j)}_{\downarrow}\vec{c}_{j+\Delta j}^{\;\rm (L)}
 \\
h^{(j,\Delta j)}_{\uparrow}\vec{c}_{j+\Delta j}^{\;\rm (R)}
\end{array}
\right)
\label{recursion2}
\end{equation}
for arbitrary $\gamma^{(\mu)}_l$, Eqs. (\ref{app-1}),
(\ref{app-2}),(\ref{app-3}) and (\ref{app-4}) 
are equivalent to
\begin{eqnarray}
h_\downarrow^{(j_{\rm l} ,1)}
\vec{c}^{\;(\rm L)}_{j_{\rm l}+1}
=
h_\downarrow^{(j_{\rm l} ,1)}\vec{c}^{\;(\rm D,\downarrow)}_{j_{\rm l}+1} 
+W^{(j_{\rm l} ,1)}\vec{c}^{\;(\rm D,\uparrow)}_{j_{\rm l}+1}
\label{app-5}
\end{eqnarray}

\begin{eqnarray}
H^{(j_{\rm l}+1 ,-1)}\vec{c}^{\;(\rm D)}_{j_{\rm l}}
=
\left(
\begin{array}{c}
h_\downarrow^{(j_{\rm l}+1 ,1)}
\\
\;^tW^{(j_{\rm l} ,1)}
\end{array}
 \right)\vec{c}^{\;(\rm L)}_{j_{\rm l}}
\label{app-6}
\end{eqnarray}

\begin{eqnarray}
H^{(j_{\rm r} ,1)}\vec{c}^{\;(\rm D)}_{j_{\rm r}+1}
=
\left(
\begin{array}{c}
W^{(j_{\rm r} ,1)} \\
\\
h_\uparrow^{(j_{\rm r} ,1)}
\end{array}
 \right)\vec{c}^{\;(\rm R)}_{j_{\rm r}+1}
\label{app-7}
\end{eqnarray}

\begin{eqnarray}
h_\uparrow^{(j_{\rm r}+1 ,1)}\vec{c}^{\;(\rm R)}_{j_{\rm r}}
=\;^tW^{(j_{\rm r} ,1)}\vec{c}^{\;(\rm D,\downarrow)}_{j_{\rm r}}
+h_\uparrow^{(j_{\rm r}+1 ,1)}\vec{c}^{\;(\rm D,\uparrow)}_{j_{\rm r}}.
\label{app-8}
\end{eqnarray}
Multiplying inverse matrices of
$ h_\downarrow^{(j_{\rm l} ,1)}
,H^{(j_{\rm l}+1 ,-1)}, H^{(j_{\rm r} ,1)}$
 and $h_\uparrow^{(j_{\rm r}+1 ,1)}$  , we can derive
 the boundary conditions (\ref{LD}) and (\ref{DR})
 from Eqs. (\ref{app-5}),(\ref{app-6}),(\ref{app-7}) and (\ref{app-8}).

In  the following formulas,
 we rewrite Eq. (\ref{f-expand}) 
as
\begin{equation}
 \left(
\begin{array}{c}
\vec{c}^{\;(\mu)}_{2m-1} \\
\vec{c}^{\;(\mu)}_{2m} \\
\end{array}
\right)
=
\left(
\begin{array}{cc}
U_{-1,+}^{(\mu)}\Lambda_{\mu}^m,&
U_{-1,-}^{(\mu)}\Lambda_{\mu}^{-m}
 \\
U_{0,+}^{(\mu)}\Lambda_{\mu}^m,&
U_{0,-}^{(\mu)}\Lambda_{\mu}^{-m}
\end{array}
\right)
\left(
\begin{array}{c}
\vec{\gamma}_+^{\;(\mu)}
 \\
\vec{\gamma}_-^{\;(\mu)}
\end{array}
\right)
\label{f-expand2}
\end{equation}
where $\Lambda_{\mu}$ 
is the diagonal matrices
 of which the diagonal element is 
$\left[\Lambda_{\mu}\right]_{l,l}=
\lambda_{l}^{(\mu)}$.
We introduce notations for region D 
 that are $\;^t\vec{\gamma}^{\;\rm (D)}
=(\;^t\vec{\gamma}_+^{\;\rm (D)},\;^t\vec{\gamma}^{\;\rm (D)}_-)$,
\begin{equation}
\left(
\begin{array}{c}
U_{\nu}^{\;\rm (D,\downarrow)} \\
U_{\nu}^{(\rm D,\uparrow)} 
\end{array}
\right)
=
\left(
\begin{array}{cc}
U_{\nu,+}^{(\rm D)},& U_{\nu,-}^{(\rm D)}
\end{array}
\right)
\end{equation}
\begin{equation}
\widetilde{\Lambda}_{\rm D}
=
\left(
\begin{array}{cc}
\Lambda_{\rm D},& 0
 \\
 0,& \Lambda_{\rm D}^{-1}
\end{array}
\right)
\end{equation}
where $\nu= -1,0$.
Using these notations, we transform
the boundary conditions (\ref{LD}) and (\ref{DR}) 
 into
\begin{equation}
\left(
\begin{array}{c}
\vec{\gamma}^{\;\rm (D)}
\\
\vec{\gamma}_-^{\;\rm (L)}
\\
\vec{\gamma}^{\;\rm (R)}_+
\end{array}
\right)
=
\widetilde{S}
\left(
\begin{array}{c}
\vec{\gamma}_+^{\;\rm (L)}
\\
\vec{\gamma}^{\;\rm (R)}_-
\end{array}
\right)
\end{equation}
where
\begin{equation}
\widetilde{S}
=
-\left(
\begin{array}{ccc}
Y_{\rm L}, & Z_{\rm L,-},& 0 
 \\
Y_{\rm R}, & 0,&  Z_{\rm R,+}
\end{array}
\right)^{-1}
\left(
\begin{array}{cc}
Z_{\rm L,+},& 0 \\
0, &  Z_{\rm R,-} 
\end{array}
\right).
\label{app-tildeS}
\end{equation}
Matrices $Y_{\rm L}$ and $Z_{\rm L,\pm} $ are defined by
\begin{equation}
Y_{\rm L} \\
=
\left(
\begin{array}{c}
-\left[ U_{-1-j_{\rm l}}^{(\rm D,\downarrow)}+
q^{\downarrow}_{j_{\rm l}}U_{-1-j_{\rm l}}^{(\rm D,\uparrow)}\right]\widetilde{\Lambda}_{\rm D}^{1+j_{\rm l}}
\\
-U_{j_{\rm l}}^{(\rm D,\downarrow)}
\\
-U_{j_{\rm l}}^{(\rm D,\uparrow)}
\end{array}
\right)
\label{YL}
\end{equation}
\begin{equation}
Z_{\rm L, \pm} \\
=
\left(
\begin{array}{c}
U_{-1-j_{\rm l},\pm}^{(\rm L)}\Lambda_{\rm L}^{\pm(1+j_{\rm l})}
\\
U_{j_{\rm l},\pm}^{(\rm L)}
\\
0
\end{array}
\right)
\label{ZL}
\end{equation}
where
\begin{eqnarray}
q^{\downarrow}_j
=\frac{1}{h_\downarrow^{(j ,1)}}W^{(j ,1)}
\end{eqnarray}
and $j_{\rm l}$ is either $-1$ or 0.
Matrices $Y_{\rm R}$ and $Z_{\rm R,\pm} $ are defined by
\begin{equation}
Y_{\rm R} \\
=
\left(
\begin{array}{c}
-U_{\Delta j_{\rm r}}^{(\rm D,\uparrow)}-q_{\Delta j_{\rm r}}^{\uparrow}U_{\Delta j_{\rm r}}^{(\rm D,\downarrow)}\\
-U_{-\Delta j_{\rm r}-1}^{(\rm D,\uparrow)}\widetilde{\Lambda}_{\rm D}^{\Delta j_{\rm r}+1}
\\
-U_{-\Delta j_{\rm r}-1}^{(\rm D,\downarrow)}\widetilde{\Lambda}_{\rm D}^{\Delta j_{\rm r}+1}
\end{array}
\right)\widetilde{\Lambda}_{\rm D}^M
\label{YR}
\end{equation}
and
\begin{equation}
Z_{\rm R,\pm} \\
=
\left(
\begin{array}{c}
U_{\Delta j_{\rm r},\pm}^{(\rm R)}
\\
U_{-\Delta j_{\rm r}-1,\pm}^{(\rm R)}\Lambda_{\rm R}^{\pm(\Delta j_{\rm r}+1)}
\\
0
\end{array}
\right)\Lambda_{\rm R}^{\pm M}
\label{ZR}
\end{equation}
where $\Delta j_{\rm r}$
 is either 0 or $-1$,
\begin{eqnarray}
q_j^{\uparrow}=
\frac{1}{h_\uparrow^{(j+1 ,1)}}
\;^tW^{(j ,1)}
\end{eqnarray}
and $M$ is the integer satisfying $j_{\rm r}=2M+\Delta j_{\rm r}
$.
The $S_{\rm RL}$ matrix (\ref{smatrix}) is derived from the 
the $\widetilde{S}$ matrix (\ref{app-tildeS})
as
$(r_{\rm LL})_{i,i'}=\widetilde{S}_{4n_{\rm D}+i,i'}$,
$(t_{\rm RL})_{j,i}=\widetilde{S}_{2n_{\rm L}+4n_{\rm D}+j,i}$,
$(t_{\rm LR})_{i,j}=\widetilde{S}_{4n_{\rm D}+i,
2n_{\rm L}+j}$
and 
$(r_{\rm RR})_{j,j'}=\widetilde{S}_{2n_{\rm L}+4n_{\rm D}+j,
2n_{\rm L}+j'}$
 where $1\leq i \leq \overline{N}_{\rm L},1\leq j \leq \overline{N}_{\rm R}$.
The numerical errors are estimated by 
\begin{equation}
\sigma_{\rm sym}=\sum_{i=1}^{N_{\rm S}}
\sum_{j=1}^{N_{\rm S}}
|(S_{\rm RL})_{i,j}-(S_{\rm RL})_{j,i}|
\end{equation}
and
\begin{equation}
\sigma_{\rm uni}=\sum_{i=1}^{N_{\rm S}}
\sum_{j=1}^{N_{\rm S}}
\left|
 \sum_{k=1}^{N_{\rm S}}
(S_{\rm RL})_{k,i}^*(S_{\rm RL})_{k,j} -\delta_{i,j})\right|
\end{equation}
where $N_{\rm S}=\overline{N}_{\rm L}+\overline{N}_{\rm R}$.
In the exact numerical calculations of Sec. \ref{sec-exact},  $N_{\rm S}=4$ and
the numerical errors are quite small
 as $\sigma_{\rm uni} < 2.2\times 10^{-11}, \sigma_{\rm sym} < 1.2\times 10^{-11}$.

\section{perturvative calculation of $S_\mu$  }
We define 2 $\times$ 4 matrices $U_{\rm L}^{[n]}$ and $U_{\rm R}^{[n]}$
as 
\begin{equation}
\left(
\begin{array}{l}
U_{\rm L}^{[n]} \\
U_{\rm R}^{[n]}
\end{array}
\right)
\equiv
(\vec{D}_{+,+}^{\;[n](-)},\;\vec{D}_{-,+}^{\;[n](+)},\;\vec{D}_{+,-}^{\;[n](-)},\;\vec{D}_{-,-}^{\;[n](+)})
\label{def-ULR}
\end{equation}
where
\begin{equation}
\;^t\vec{D}_{\sigma,\tau}^{\;[0](\zeta)}
=\frac{1}{2}(1,\sigma,\tau f_\sigma^{(\zeta)}
, \tau\sigma f_\sigma^{(\zeta)})
\label{def-dzero2}
\end{equation}
 and $\vec{D}_{\sigma,\tau}^{\;[1](\zeta)}$ 
 is defined by Eq. (\ref{perturb-wave}) of which
 $\vec{b}_{-\sigma,\tau'}^{\;[0](\zeta)}$
 is replaced by $\vec{D}^{\;[0](\zeta)}_{-\sigma,\tau'}$.
With this definition, Eq. (\ref{cj-Un}) is rewritten as
\begin{equation}
U^{[n]}_{\rm D } =
\left(
\begin{array}{c}
\frac{1}{\sqrt{2 n_\downarrow}}
U_{\rm L}^{[n]}
\\
\frac{1}{\sqrt{2 n_\uparrow}}
U_{\rm R}^{[n]}
\end{array}
\right)
\label{appendix-cj-D2}
\end{equation}

In contrast to the exact calculation,
boundary conditions (\ref{LD}) and (\ref{DR})
 are approximated by
\begin{eqnarray}
\left(
\begin{array}{c}
\vec{c}^{\;(\rm L)}_{j_{\rm l}+1} 
\\
\vec{c}^{\;(\rm L)}_{j_{\rm l}}
\\
0
\end{array}
\right)
=
\left(
\begin{array}{c}
\vec{c}^{\;(\rm D,\downarrow)}_{j_{\rm l}+1} 
\\
\vec{c}^{\;(\rm D,\downarrow)}_{j_{\rm l}}
\\
\vec{c}^{\;(\rm D,\uparrow)}_{j_{\rm l}}
\end{array}
\right)
\label{appendix-LD}
\end{eqnarray}
and
\begin{eqnarray}
\left(
\begin{array}{c}
\vec{c}^{\;(\rm R)}_{j_{\rm r}}
\\
\vec{c}^{\;(\rm R)}_{j_{\rm r}+1}
\\
0 
\end{array}
 \right)
=
\left(
\begin{array}{c}
\vec{c}^{\;(\rm D,\uparrow)}_{j_{\rm r}}
\\
\vec{c}^{\;(\rm D,\uparrow)}_{j_{\rm r}+1}
\\
\vec{c}^{\;(\rm D,\downarrow)}_{j_{\rm r}+1}
 \\
\end{array}
 \right)
\label{appendix-DR}
\end{eqnarray}
 in the perturbation calculation.
We derive
 matrix $X_\xi^{[n]}$ of Eq. (\ref{scatter-diamond}) 
 from Eqs. (\ref{cj-D2}),(\ref{appendix2-cj}),(\ref{appendix-cj-D2}),(\ref{appendix-LD}) 
and (\ref{appendix-DR}) as
\begin{equation}
X_{\mu}^{[n]}
=\left(
\begin{array}{cc}
U_\mu^{[n]}\Xi, & -\sqrt{2}v_0\Omega_0^*\delta_{n,0}\\
U_\mu^{[n]}, & -\sqrt{2}v_0\delta_{n,0}\\
U_{-\mu}^{[n]}, & 0\\
\end{array}
\right)
\label{def-X}
\end{equation}
where $\mu$ and $-\mu$ are complementary as
$(\mu,-\mu)=$ (L,R), (R,L) and
 $v_0=(\sigma_x+\sigma_z)/2$  with Pauli matrices (\ref{pauli}).
Under the conditions  $|w_{\sigma,\sigma}| \ll t$ 
 and $|E|  \ll t$,
 we approximate  $\Omega \simeq {\bf 1}$  and
 $\Omega_0 \simeq \widetilde{\Omega}_0$  
 where 
\begin{equation}
\widetilde{\Omega}_0
=\left(
\begin{array}{cc}
e^{i\frac{2}{3}\pi},& 0 \\
0,& e^{-i\frac{2}{3}\pi}
\end{array}
\right).
\label{omega0-2}
\end{equation}
Using this approximation in Eq. (\ref{def-X}), we show
\begin{equation}
X^{[0]}_{\rm L}
=\left(
\begin{array}{ccc}
v_0\widetilde{\Omega}_0,&  v_0\widetilde{\Omega}_0,
 & -\sqrt{2}v_0\widetilde{\Omega}_0^*\\
v_0,&  v_0
, & -\sqrt{2}v_0\\
 v_0F,& -v_0F,
 & 0\\
\end{array}
\right)
\label{X0}
\end{equation}

\begin{eqnarray}
X^{[1]}_{\rm L}
&=& 
\frac{2}{E}\left(
\begin{array}{ccc}
v_1\widetilde{\Omega}_0,& -v_1\widetilde{\Omega}_0,
 & 0\\
v_1,& -v_1
, & 0\\
v_2
,& v_2
 & 0\\
\end{array}
\right)
\label{X1L}
\end{eqnarray}

\begin{equation}
X^{[0]}_{\rm R}
=\left(
\begin{array}{ccc}
v_0F^*\widetilde{\Omega}_0,& - v_0F^*\widetilde{\Omega}_0,
 & -\sqrt{2}v_0\widetilde{\Omega}_0^*\\
v_0F^*,& - v_0F^*
, & -\sqrt{2}v_0\\
 v_0,&  v_0,
 & 0\\
\end{array}
\right)
\label{X0R}
\end{equation}

\begin{eqnarray}
X^{[1]}_{\rm R}
&=& 
\frac{2}{E}\left(
\begin{array}{ccc}
v_2^*\widetilde{\Omega}_0,& v_2^*\widetilde{\Omega}_0,
 & 0\\
v_2^*,& v_2^*
, & 0\\
v_1^*
,&- v_1^*
 & 0\\
\end{array}
\right)
\label{X1R}
\end{eqnarray}

where $v_1=\frac{1}{4}(i\sigma_y+{\bf 1}_2)G^*F$ 
 and  $v_2=\frac{1}{2}\sigma_z v_0\sigma_xG^*\sigma_x$.

Inverse of Eq. (\ref{X0}) is represented by
\begin{equation}
\left(X^{[0]}_{\rm L}\right)^{-1}
=\left(
\begin{array}{ccc}
-v_3,& \widetilde{\Omega}_0^*v_3,
 & F^*v_0\\
-v_3,& \widetilde{\Omega}_0^*v_3
, & -F^*v_0 \\
 -\sqrt{2}v_3,& \sqrt{2} \widetilde{\Omega}_0 v_3,
 & 0\\
\end{array}
\right)
\label{inverse-X0}
\end{equation}
\begin{equation}
\left(X^{[0]}_{\rm R}\right)^{-1}
=\left(
\begin{array}{ccc}
-v_4,& \widetilde{\Omega}_0^*v_4,
 & v_0\\
v_4,& -\widetilde{\Omega}_0^*v_4
, & v_0 \\
 -\sqrt{2}F^*v_4,& \sqrt{2} F^*\widetilde{\Omega}_0 v_4,
 & 0\\
\end{array}
\right)
\label{inverse-X0R}
\end{equation}
where $v_3= \sqrt{3}(i{ \bf 1}_2-\sigma_y)/6$, 
 and $v_4=i\sigma_zFv_0/\sqrt{3}$.
Using Eqs. (\ref{XS-0}),
(\ref{X0}),(\ref{X1L}),(\ref{X0R}),(\ref{X1R}),
(\ref{inverse-X0})  and 
(\ref{inverse-X0R}),
 we obtain $S_\mu^{[0]}$ and $S_\mu^{[1]}$.
 Because $^tS_{\mu}=S_{\mu}$  and  $S_\mu^*S_{\mu}=\bf{1}$ 
(see  Appendix A),
\begin{equation}
^tS_\mu^{[n]}=S_\mu^{[n]} 
\label{unitary-sym}
\end{equation}
\begin{equation}
S_\mu^{[0]*}S_\mu^{[0]}=\bf{1}
\label{unitary-0}
\end{equation}
and 
\begin{equation}
S_\mu^{[1]*}S_\mu^{[0]}+S_\mu^{[0]*}S_\mu^{[1]}=0.
\label{unitary-1}
\end{equation}
We can easily confirm that
 $S_\mu^{[0]}$ and $S_\mu^{[1]}$
 of Sec. \ref{sec-appro} satisfy Eqs. (\ref{unitary-sym}) ,(\ref{unitary-0}) 
and  (\ref{unitary-1}).

\begin{table}
\begin{tabular}{|c|c|c|c|c|
} 
 & $w_{+,+}$  & $w_{-,-}$  & $w_{-,+}$  & $w_{+,-}$  \\
 Fig. 5 & 7.7$\times 10^{-3}$ & $-9.5\times 10^{-3}$ & 
$-8.4 \times 10^{-3}$ & 
$9.5\times 10^{-3}$ \\
 Fig. 6 & 9.4$\times 10^{-2}$ & 0 & 
$2.0 \times 10^{-2}$ & 
0 \\
\end{tabular}
\caption{
The parameters defined by Eq. (\ref{overline-w})  
 for the junctions of Figs. 5 and 6
 in units of eV.
}
\end{table}

\begin{figure}
  \epsfxsize=\columnwidth
\centerline{\hbox{   \epsffile{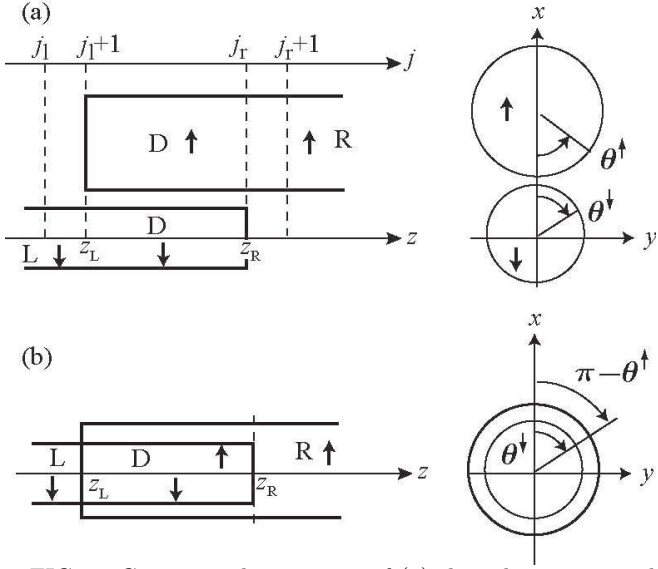} }}
\caption{Geometrical structures of (a) the side contact
 and (b) the telescoped coaxial contact.
The  single wall armchair NTs are denoted
 by $\downarrow$ and $\uparrow$.
The $z$ axis is chosen to be the axis of tube $\downarrow$.
The atomic $z$ coordinates in tubes  $\downarrow$ and $\uparrow$
 are $aj/2$ and $aj/2+\Delta z$, respectively, with integers $j$, the lattice constant $a=0.246$ nm and a small translation $|\Delta z| < a/4$.
Tubes $\downarrow$ and $\uparrow$ have
the open edges 
at $z_{\rm R}=aj_{\rm r}/2$
 and  $z_{\rm L}=\Delta z+ a(j_{\rm l}+1)/2$, respectively.
The geometrical overlap length is $z_{\rm R}-z_{\rm L}$
 while the integer overlap length $N$ is 
 defined as $N=j_{\rm r}-j_{\rm l}+1=2+2(z_{\rm R}-z_{\rm L}+ \Delta z)/a$.
 Without losing generality, $j_{\rm l}=-1, 0$.
 }
\end{figure}

\newpage

\begin{figure}
  \epsfxsize=\columnwidth
\centerline{\hbox{   \epsffile{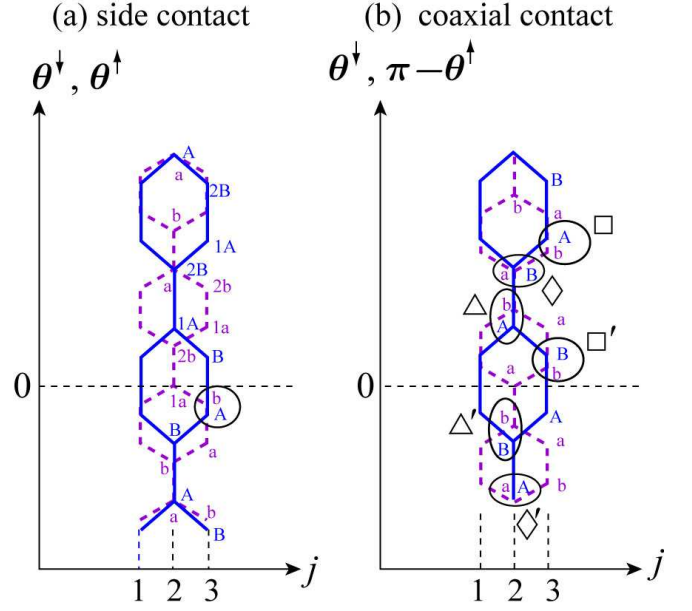} }}
\caption{
Interlayer configuration of (a) the side contact and (b) the coaxial contact
 for the case
 where $(n_\downarrow, n_\uparrow)=
(10,15)$  and $(\Delta\theta,\Delta z)=(0,0)$.
 }
\end{figure}

\newpage

\begin{figure}
  \epsfxsize=\columnwidth
\centerline{\hbox{   \epsffile{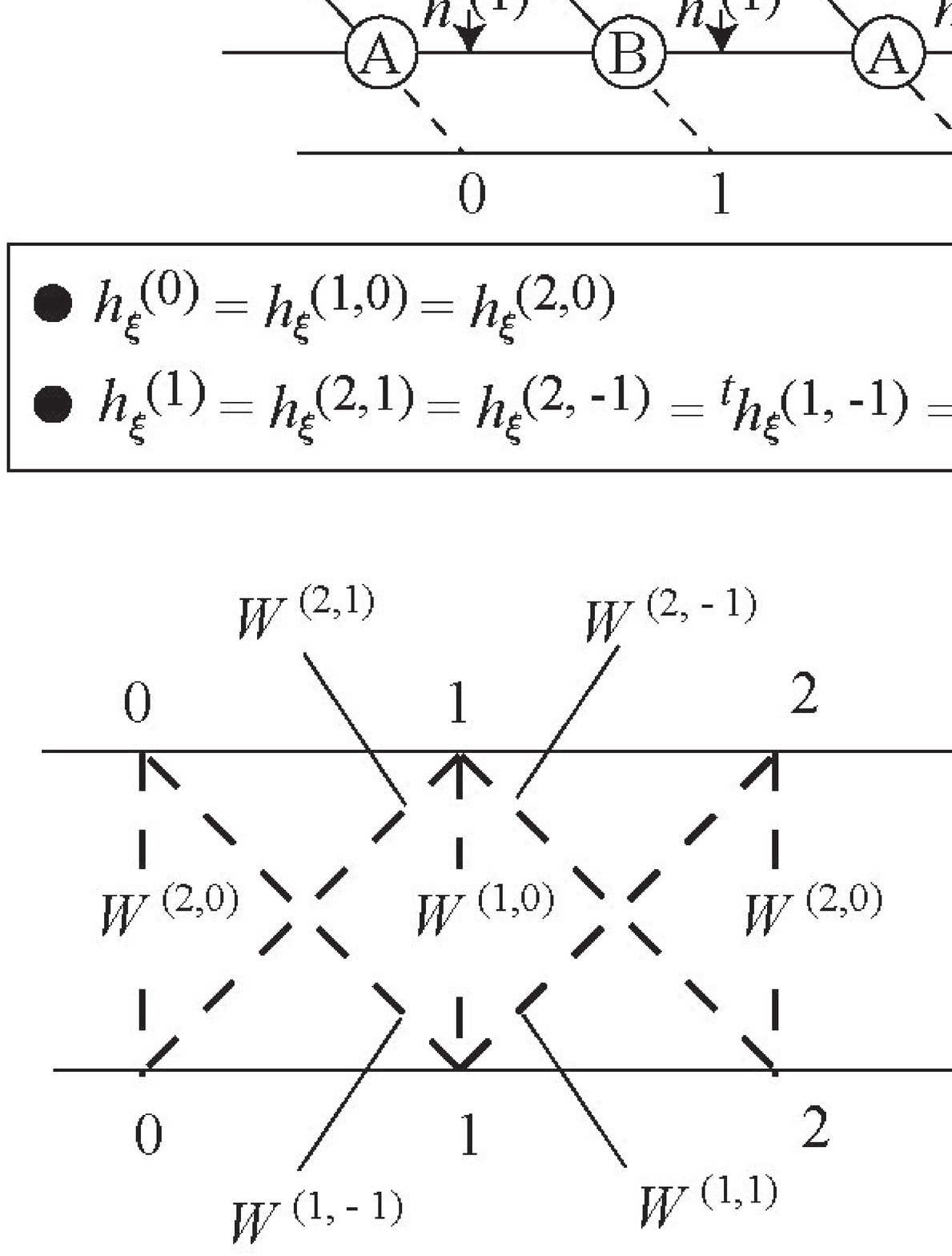} }}
\caption{ Schematic diagram  of the tight bindding Hamiltonian.
  Since $h_\xi^{(1,0)}=h_\xi^{(2,0)}$ and $h_\xi^{(j,1)}=h_\xi^{(j,-1)}$, 
  we use the abbreviations $h_\xi^{(0)}$ and $h_\xi^{(1)}$.}
\end{figure}

\newpage

\begin{figure}
  \epsfxsize=\columnwidth
\centerline{\hbox{   \epsffile{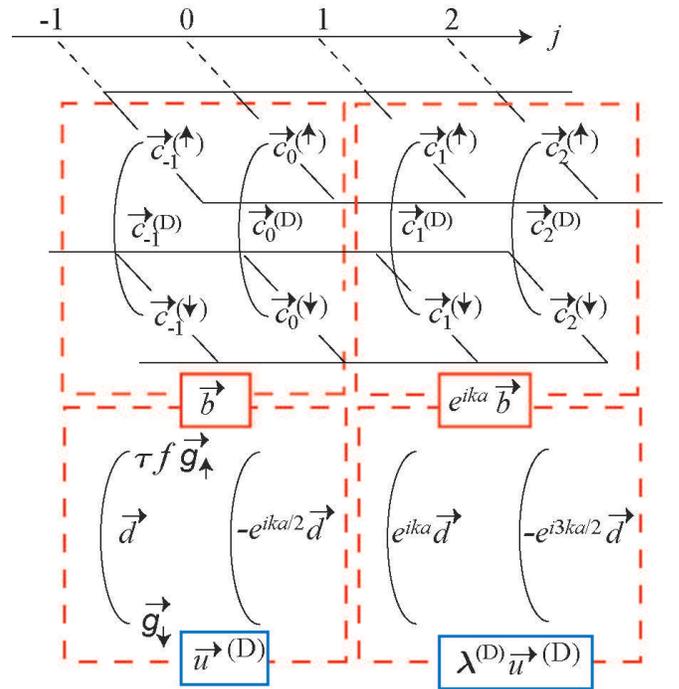} }}
\caption{
 Relation between Sec. III A and Sec. III B in notation
 of the vectors. }
\end{figure}

\newpage

\begin{figure}
  \epsfxsize=0.9\columnwidth
\centerline{\hbox{   \epsffile{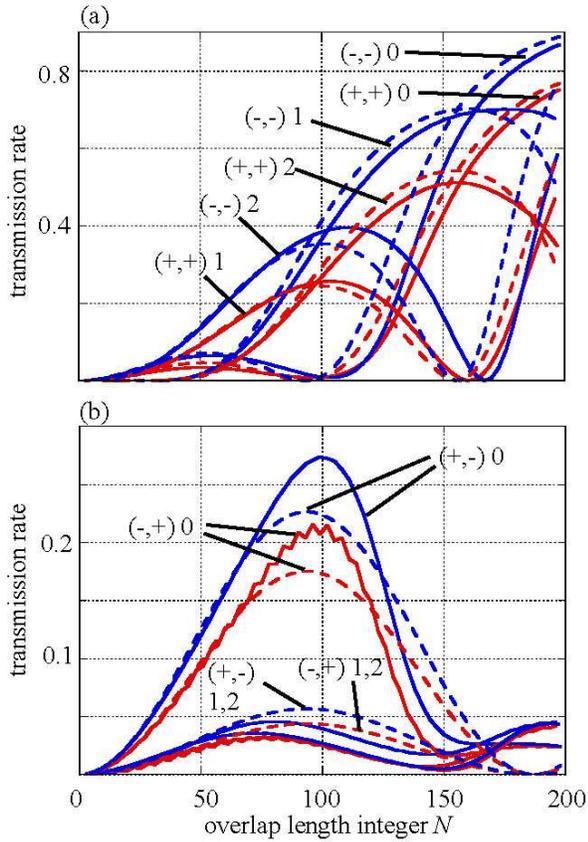} }}
\caption{
(a) Diagonal  $T_{\sigma,\sigma}$ and (b) off-diagonal  $T_{-\sigma,\sigma}$
 transmission rate  of the side contact 
$(n_\downarrow,n_\uparrow)=(10,15), j_{\rm l}=-1, 
 \Delta \theta=0,
\Delta z =0$ 
with the energy $E= $ 0.08 eV.
 The horizontal axis is  the integer $N$.
 The geometrical overlapped length equals  $(N-2)a/2$
 as is shown by Fig. 1.
Solid and dashed lines represent the exact results
and the approximate formulas, respectively.
By the attached symbols, subscripts 
of $T_{\sigma',\sigma}$ and  integers mod($N,3$) are indicated.
 The labels '$(\pm,\mp)$ 1' and '$(\pm,\mp)$ 2'  are not displayed 
 for the solid lines in (b).
 Among the four solid lines without the labels,
 that of $(+,-)$ 1
 is slightly larger 
 than the others.
 }
\end{figure}

\newpage

\begin{figure}
  \epsfxsize=0.9\columnwidth
\centerline{\hbox{   \epsffile{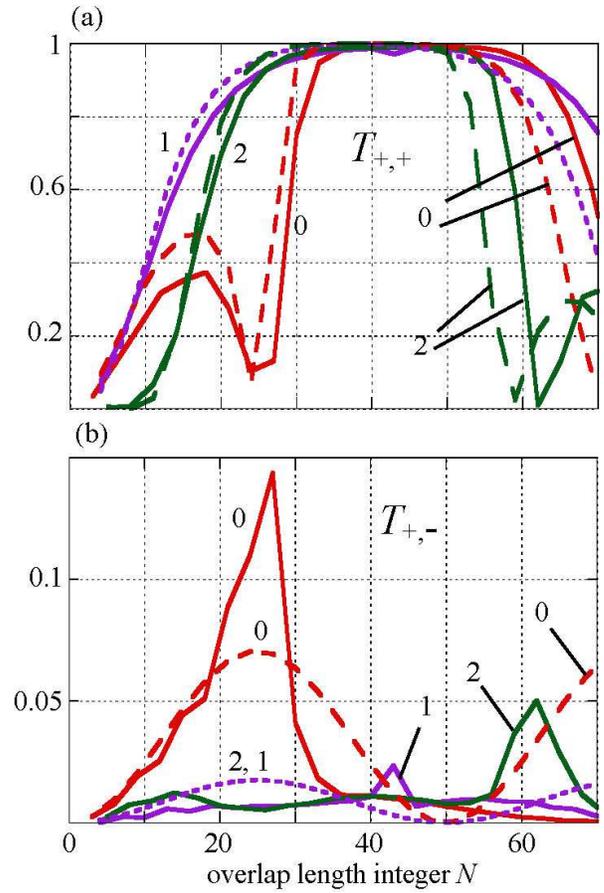} }}
\caption{
Transmission rates (a) $T_{+,+}$  and (b) $T_{+,-}$  of the coaxial contact 
$(n_\downarrow,n_\uparrow)=(10,15), j_{\rm l}=-1, \Delta\theta=0, \Delta z=0$
 with the energy $E= $ 0.30 eV.
 The horizontal axis is  the integer $N$. 
Solid and dashed lines represent the exact results
and the approximate formulas, respectively.
The attached integers 0, 1 and 2 represent mod($N,3$).
 }
\end{figure}

\newpage

\begin{figure}
  \epsfxsize=\columnwidth
\centerline{\hbox{   \epsffile{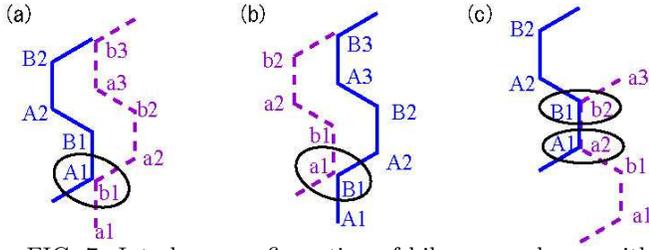} }}
\caption{Interlayer configuration of
 bilayer graphenes with (a) Ab , (b) Ba  and (c) Aa stacking.
Here (A,B) and (a,b) denote sublattices in lower $\downarrow$ 
 and upper $\uparrow$  layers, respectively.
 }
\end{figure}

\begin{figure}
  \epsfxsize=0.8\columnwidth
\centerline{\hbox{   \epsffile{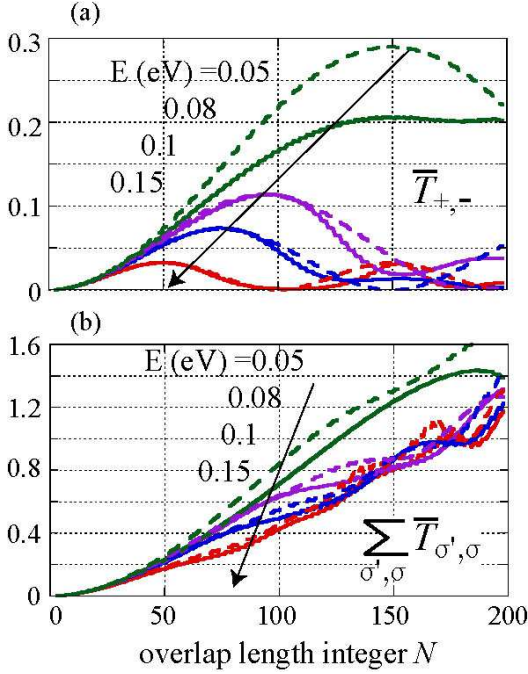} }}
\caption{
(a) $\overline{T}_{+,-}$ 
 and (b) $\sum_{\sigma' =\pm} \sum_{\sigma=\pm} \overline{T}_{\sigma',\sigma}$ 
for the energies $E=0.05, 0.08, 0.1$ and 0.15 eV.
Solid and dashed lines represent the exact results
and the approximate formulas, respectively.
 Here  $\overline{T}_{\sigma',\sigma}(N)$  denotes
 the smoothed transmission rate 
of the junction of Fig. 5
  defined by $\frac{1}{3} \sum_{j=-1}^1T_{\sigma',\sigma}(N+j)$.
 }
\end{figure}

\newpage

\begin{figure}
  \epsfxsize=0.7\columnwidth
\centerline{\hbox{   \epsffile{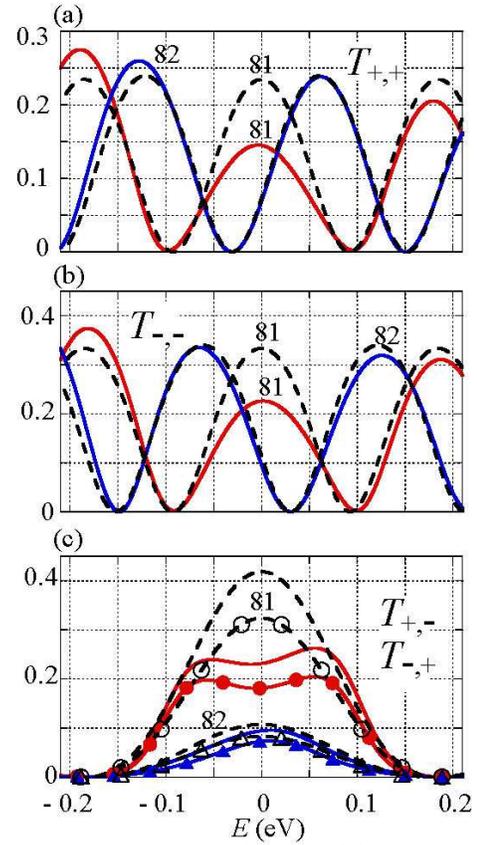} }}
\caption{Transmission rates (a) $T_{+,+}$ (b) $T_{-,-}$ 
and (c) $T_{\pm,\mp}$ of the junction of Fig. 5
 as a function energy $E$
 when $N=81, 82$.
Solid and dashed lines represent the exact results
and the approximate formulas, respectively.
In (c), solid lines with closed symbols
 and dashed lines with open symbols correspond to 
 $T_{-,+}$.
 }
\end{figure}

\begin{figure}
  \epsfxsize=\columnwidth
\centerline{\hbox{   \epsffile{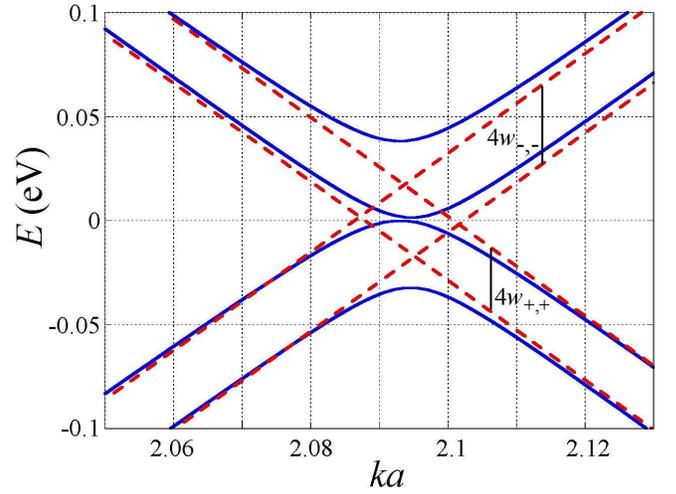} }}
\caption{The dispersion relation corresponding to region D of the junction of
 Fig. 5. 
Solid and dashed lines represent the exact results
and the approximate formulas (\ref{linear-k}), respectively.
 }
\end{figure}
\newpage

\begin{figure}
  \epsfxsize=\columnwidth
\centerline{\hbox{   \epsffile{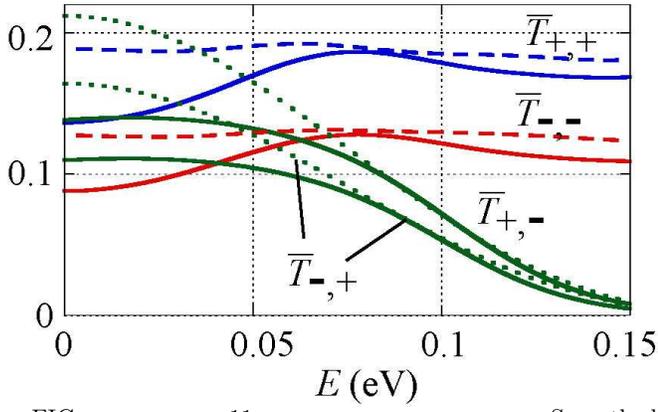} }}
\caption{
Smoothed transmission rates $\frac{1}{3} \sum_{j=-1}^1T_{\sigma',\sigma}(N+j)
$ as a function of the
 energy $E$ for the junction of Fig. 5 when $N=82$.
Solid and dashed lines represent the exact results
and the approximate formulas, respectively.
 }
\end{figure}

\begin{figure}
  \epsfxsize=\columnwidth
\centerline{\hbox{   \epsffile{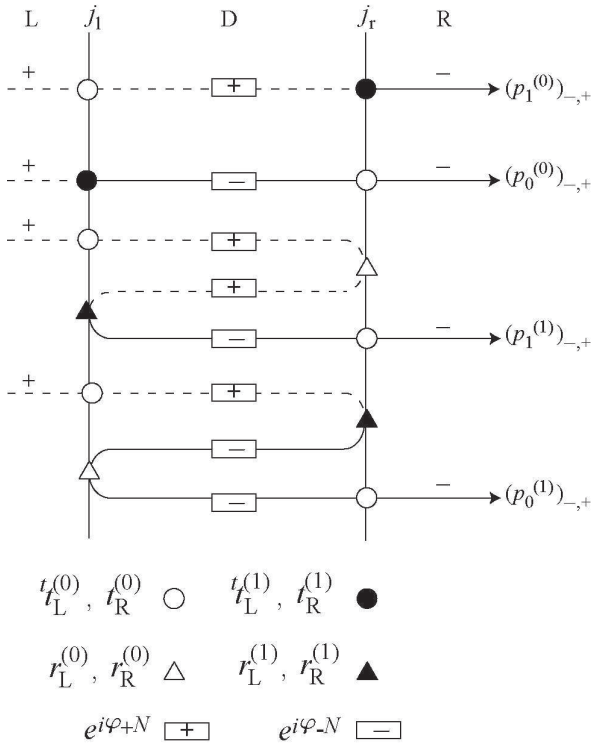} }}
\caption{
Multiple reflection  with the notation
 of Eq. (\ref{multiple-scatter2}) in the case where the symmetric (+) channel  is incident
 from region L.
 }
\end{figure}

\newpage

\begin{figure}
  \epsfxsize=\columnwidth
\centerline{\hbox{   \epsffile{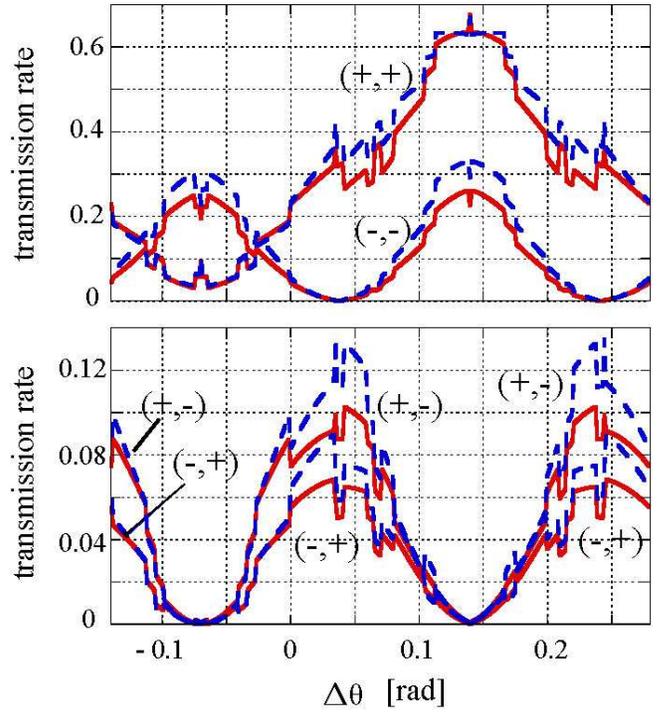} }}
\caption{
Transmission rate $T_{\sigma',\sigma}$ 
 as a function of $\Delta \theta$  
in the case where $(n_\downarrow,n_\uparrow) =(10,15), N=82, \Delta z=0, j_{\rm l}=-1$ and $E=0.05$ eV.
Solid and dashed lines represent the exact results
and the approximate formulas, respectively.
The data are limited to  the discrete 
 $\Delta \theta =m\pi/(150n_\uparrow) $ 
 with intejers $m$.
 }
\end{figure}

\begin{figure}
  \epsfxsize=\columnwidth
\centerline{\hbox{   \epsffile{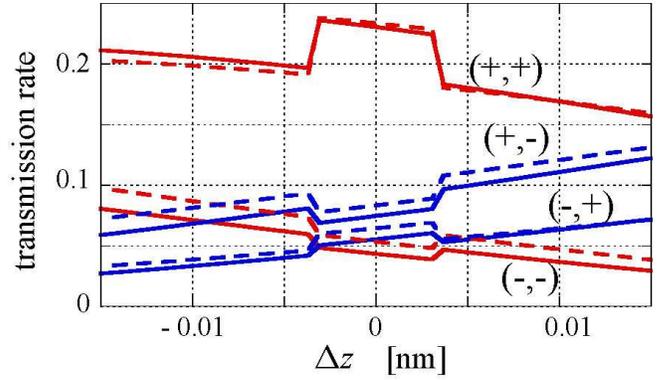} }}
\caption{
The transmission rate $T_{\sigma',\sigma}$ 
 as a function of $\Delta z$ in the case where $(n_\downarrow,n_\uparrow) =(10,15), N=82 , \Delta\theta=0, j_{\rm l}=-1$   and $E=0.05$ eV.
Solid and dashed lines represent the exact results
and the approximate formulas, respectively.
The data are limited to  the discrete 
 $\Delta z =ma/400$ 
 with intejers $m$.
 }
\end{figure}


\newpage

\begin{thebibliography}{10}
\bibitem{NT-book} 
R.  Saito, G.  Dresselhaus, and M.  S.  Dresselhaus, {\it Physical Propertiesof
  Carbon Nanotubes} (Imperial College Press, London,1998). 
\bibitem{NT-review} 
J.-C.  Charlier, X.  Blase, and S.  Roche, 
Rev.  Mod.  Phys.  {\bf 79}, 677 (2007). 
\bibitem{graphene-review} 
S. D. Sarma, S. Adam, E. H. Hwang, and E.  Rossi, 
Rev.  Mod.  Phys.  {\bf 83}, 407 (2011). 



\bibitem{pseudo-gap}
Y.-K.  Kwon, S. Saito, and D.  Tom\'{a}nek,
 Phys.  Rev.  B {\bf 58},  13314(R)  (1998).;Y.-K.  Kwon, and D.  Tom\'{a}nek,
 Phys.  Rev.  B {\bf 58}, 16001(R) (1998); 
Y.   Miyamoto, S.  Saito, and D.  Tom\'{a}nek, {\it ibid.} {\bf 65}, 041402 (2001). 

\bibitem{3-NFE}
S. Okada, A. Oshiyama, and S. Saito, Phys.  Rev.  B {\bf 62},  7634  (2000).


\bibitem{4-rope}
J. Tersoff and R. S. Ruoff, Phys.  Rev.  Lett.  {\bf 73}, 676 (1994).


\bibitem{bilayer-E-fierld}
H. M. Abdullah, M. A.  Ezzi, and H. Bahlouli, J. of Appl. Phys. {\bf 124}, 204303 (2018);T.S. Li, Y.C. Huang, S.C. Chang, Y.C. Chuang, and M.F. Lin, Eur. Phys. J. B 
{\bf 64}, 73 (2008).


\bibitem{6-twist}
M. Ochi, M. Koshino, and K. Kuroki, Phys.  Rev.  B {\bf 98},  081102(R)  (2018);
Y. Cao, V. Fatemi, S. Fang, K. Watanabe, T. Taniguchi, E. Kaxiras, and P. J-Herrero, Nature(London) {\bf 556} 43 (2018).


\bibitem{17-mono-double-junction}
T.  Nakanishi, M.  Koshino, and T.  Ando, Phys.  Rev.   B {\bf 82}, 
 125428 (2010); M. Koshino {\it ibid} {\bf 88}, 115409 (2013).


\bibitem{experiment-telescope-mechanical} 
J.  Cumings and A.  Zettl, Science {\bf 289}, 602 (2000); A.  Kis, K.  Jensen,S.  Aloni, W.  Mickelson, and A.  Zettl, Phys.  Rev.  Lett.  {\bf  97}, 025501 (2006);
S.  Akita and Y.  Nakayama, J.  J.  Appl.  Phys. {\bf 42}, 4830  (2003);
M.  Nakajima, S.  Arai, Y.  Saito, F.  Arai, and T.  Fukuda, 
{\it ibid.} {\bf 46}, L1035 (2007); 
 W.  Zhang, Z.  Xi, G.  Zhang, C.  Li, and D.  Guo, Phys.  Chem.  Lett.  {\bf 112}, 14714 (2008). 


\bibitem{theory-telescope-mechanical} 
J.  Servantie and P.  Gaspard, Phys.  Rev.  B {\bf 73}, 125428 (2006); Phys.  Rev. 
  Lett.  {\bf 91}, 185503 (2003);
Q.  Zheng and Q.  Jiang, Phys.  Rev.  Lett.  {\bf 88}, 045503 (2002);
  S.  B.  Legoas,
  V.  R.  Coluci, S.  F.  Braga, P.  Z.  Coura, S. O.  Dantas, and D.  S.  Galvao, {\it  ibid. } {\bf 90}, 055504 (2003);
  W.  Guo, Y.  Guo, H.  Gao, Q.  Zheng, and W.  Zhong. 
  {\it ibid.} {\bf 91}, 125501 (2003); 
P.  Tangney, M.  L.  Cohen, and S.  G.  Louie, {\it ibid. } {\bf 97}, 195901
  (2006);
Q.  Zheng, J.  Z.  Liu, and Q.  Jiang, Phys. 
  Rev.  B {\bf 65}, 245409 (2002);
J. W.  Kang and O. K.  Kwon  Appl.  Sur.  Sci.  {\bf 258}, 2014 (2012).

\bibitem{gra-gra-MD} 
 A. M. Popov, I. V. Lebedeva, A. A. Knizhnik, Y. E. Lozovik, and B. V. Potapkin,
Phys. Rev. B {\bf 84},  245437  (2011).


\bibitem{NT-graphene-force-theory}
A.  Buldum and J.  P.  Lu, Phys.  Rev.  Lett., {\bf 83}, 5050 (1999);
M. Seydou, Y. J. Dappe, S. Marsaudon, J.-P. Aim\'{e}, X.  Bouju, and A.-M. Bonnot, Phys. Rev. B {\bf 83}, 045410 (2011)
; M. Seydou, S. Marsaudon, J. Buchoux, and J. P. Aim\'{e}m
 {\it ibid} {\bf 80}, 245421 (2009).

\bibitem{gra-NT-AB}
\'{A}. Szabados, L. P. Bir\'{o}, and P. R. Surj\'{a}n, Phys.  Rev.  B {\bf 73},  195404  (2006).


\bibitem{JJ-Sakurai}
J. J. Sakurai, {\it Modern quantum mechanics} (Addison-Wesley, Tokyo, 1994).




\bibitem{gra-band}
M. Koshino and T. Ando, Phys.  Rev.  B {\bf 76},  085425  (2007);
J. Nilsson, A. H. C. Neto, F. Guinea and N. M. R. Peres,
{\it ibid} {\bf 78}, 045405 (2008);
J. Ruseckas, G. Juzeliunas and I. V. Zozoulenko, {\it ibid}
B {\bf 83}, 035403 (2011); F. Zhang, Bhagawan Sahu, H. Min and A. H. MacDonald, {\it ibid} {\bf 82}, 035409 (2010)
B. Partoens and F. M. Peeters Phys. Rev. B {\bf 74}, 075404 (2006); {\bf 75},  193402  (2007).



\bibitem{10-covalent} 
J-L. Zhu, F-F. Xu and Y-F. Jia, Phys. Rev. B {\bf 74}, 155430 (2006);
M. Terrones, F. Banhart, N. Grobert, J.-C. Charlier, H. Terrones and P. M. Ajayan, Phys. Rev. Lett. {\bf 89}, 075505 (2002);
F. Y. Meng,S. Q. Shi, D. S. Xu and R. Yang, Phys. Rev. B {\bf 70}, 125418 (2004);
A. V. Krasheninnikov, K. Nordlund, J. Keinonen and F. Banhart,
 {\it ibid} {\bf  66}, 245403 (2002);
S. Dag, R. T. Senger, and S. Ciraci, {\it ibid} {\bf 70}, 205407 (2004).

\bibitem{gra-gra-10-LDL}
D, Valencia, J-Q, Lu, J. Wu, F. Liu, F. Zhai, and Y-J. Jiang
,AIP Advances {\bf 3}, 102125 (2013);
J.  Nilsson, A. H. Castro Neto, F.  Guinea and N.  M.  R.  Peres, Phys.  Rev.  B {\bf 76}, 165416 (2007).

\bibitem{experiment-telescope-conductance} 
J.  Cumings and A.  Zettl,  Phys.  Rev.  Lett.  {\bf  93}, 086801 (2004); S.  Akita and Y.  Nakayama, J.  J.  Appl.  Phys.  {\bf 43}, 3796 (2004). 


\bibitem{gra-gra-tele-cond}
 D. Yin, W. Liu , X. Li, L. Geng, X. Wang and P. Huai,
Appl. Phys. Lett. {\bf 103}, 173519 (2013);
J.  W.  Gonzalez, H.  Santos,  M.  Pacheco,  L.  Chico  and  L.  Brey, Phys.  Rev.  B {\bf 81}, 195406 (2010);
J. Zheng, P. Guo, Z. Ren, Z. Jiang, J. Bai, and Z. Zhang, Appl. Phys. Lett. {\bf 101}, 083101 (2012);
X-G. Li, I-H. Chu, X.-G. Zhang and H-P. Cheng, Phys. Rev. B {\bf 91}, 195442 (2015);
H. M. Abdullah, B. V. Duppen, M. Zarenia, H. Bahlouli,
and F. M. Peeters, J. Phys. Condens. Matter {\bf 29}, 425303 (2017);
I. V. Lebedeva , A. M. Popov , A. A. Knizhnik, Y. E. Lozovik,
 N. A. Poklonski, A. I. Siahlo, S. A. Vyrko, S. V. Ratkevich, Comp. Mat. Sci. {\bf 109} 240 (2015) .




\bibitem{15-gra-NT}
B. G. Cook, W. R. French, and K.  Varga,  Appl. Phys. Lett. {\bf 101}, 153501 (2012).


\bibitem{theory-telescope-conductance} 
Q.  Yan, G.  Zhou, S.  Hao, J.  Wu, and W.  Duan, Appl.  Phys.  Lett.  {\bf 88},
  173107  (2006);
A.  Hansson and S.  Stafstrom, Phys.  Rev.  B, {\bf 67}, 075406 (2003);
I.  M.  Grace, S.  W.  Bailey, and C.  J.  Lambert, Phys.  Rev.  B, {\bf 70}, 153405  (2004);Y.-J.  Kang, K. J.  Chang, and Y.-H.  Kim, Phys.  Rev.  B {\bf 76},  205441  (2007).

\bibitem{previous-paper}
R.  Tamura, Phys.  Rev.  B {\bf 82},  035415  (2010); { \bf 86}, 205416 (2012).

\bibitem{ref-2} 
D.-H.  Kim and K.  J.  Chang, Phys.  Rev.  B, {\bf 66}, 155402 (2002).

\bibitem{sawai} 
R.  Tamura, Y.  Sawai, and J.  Haruyama,
 Phys.  Rev.  B {\bf 72},  045413  (2005). 

\bibitem{Uryu} 
S.  Uryu and T.  Ando, Phys.  Rev.  B {\bf 76}, 155434 (2007);{\bf 72}, 245403 (2005).



\bibitem{10-side-contact} 
 S. Tripathy and T. K. Bhattacharyya, Physica E {\bf 83}, 314 (2016);
Q. Liu, G. Luo,  R. Qin, H. Li, X. Yan, C. Xu, L. Lai, J.  Zhou, S. Hou,
 E. Wang, Z. Gao  and J. Lu, Phys. Rev. B {\bf 83}, 155442 (2011);
A. Buldum and J. P. Lu,Phys. Rev. B {\bf 63}, 161403(R) (2001).


\bibitem{10-side-contact-ref-1} 
F. Xu, A.  Sadrzadeh, Zhiping Xu  and B. I. Yakobson,
J.  Appl. Phys. {\bf 114}, 063714 (2013).



\bibitem{side-tele} 
C.  Buia, A.  Buldum, and J.  P.  Lu, Phys.  Rev.  B, {\bf 67}, 113409 (2003).

\bibitem{side-tele-perturb} 
M. A.  Tunney and N. R.  Cooper, Phys.  Rev.  B {\bf 74},  075406  (2006).

\bibitem{Datta} 
S.  Datta, {\it Electronic Transport in Mesoscopic Systems} (Cambridge
  University Press, Cambridge 1995). 

\bibitem{X-junction-cond}
T. Nakanishi and T. Ando, 
J. Phys. Soc. Jpn {\bf 70}, 1647 (2001);
Y-G. Yoon, M. S. C. Mazzoni, H. J. Choi, J.  Ihm, and S. G. Louie, Phys. Rev. Lett {\bf 86} 688 (2001);A. A. Maarouf and E. J. Mele, Phys.  Rev.  B {\bf 83 }, 045402  (2011); B. G. Cook, P. Dignard, and K.  Varga,
 Phys.  Rev.  B {\bf 83}, 205105 (2011).

\bibitem{Lambin} Ph.  Lambin,  V.  Meunier and A.  Rubio,  
 Phys.  Rev.  B {\bf 62} 5129 (2000);J. -C.  Charlier,  
 J.  -P.  Michenaud and Ph.  Lambin,  {\it ibid} {\bf 46} 4540 (1992). 


\bibitem{footnote}
Single valued $t_1$ $(t_1=0.36 $ eV) of the present work
 and multivalued $t_1$  $(t_1=$ 0.36 eV, 0.16 eV)  of Ref. \protect\cite{sawai}
  show that $w_{\pm,-} = 0$ and $w_{\pm,-} \neq 0$, respectively, for the coaxial contact
 under the conditions  mod($n_\uparrow,3) =0$ and $|n_\uparrow-n_\downarrow|=5$ . This difference  is explained with the term  'three fold cancellation' in Ref. \protect\cite{sawai}.



\end{thebibliography}
\end{document}